\documentclass{ar2e}
\usepackage{axodraw}
\usepackage{epsfig}
\begin{document}
\input epsf.tex    

\input psfig.sty

\newcommand{\ifm}[1]{\relax\ifmmode #1\else $#1$\hskip 0.15cm\fi}
\newcommand{\gev}{\ifm{\mbox{ GeV}}}
\newcommand{\ttb}{\ifm{t\tbar}}
\newcommand{\mean}[1]{\ifm{{\mbox{$\bigl\langle #1 \bigr\rangle$}}}}
\newcommand{\MET}{\ifm{\mbox{$\rlap{\kern0.1em/}\ET$}}}
\newcommand{\MissingB}[2]{\ifm{\mbox{$#1\kern-0.57em\raise0.19ex\hbox{/}_{#2}$}}}
\newcommand{\MissingS}[2]{\ifm{\mbox{$#1\kern-0.42em\raise0.19ex\hbox{/}_{#2}$}}}
\newcommand{\vMissingB}[2]{\ifm{\mbox{$\vec{#1}\kern-0.57em\raise0.19ex\hbox{/}_{#2}$}}}
\newcommand{\vMissingS}[2]{\ifm{\mbox{$\vec{#1}\kern-0.42em\raise0.19ex\hbox{/}_{#2}$}}}
\newcommand{\mET}{\MissingB{E}{T}}
\newcommand{\mpT}{\MissingS{p}{T}}
\newcommand{\vmET}{\vMissingB{E}{T}}
\newcommand{\vmpT}{\vMissingS{p}{T}}

\newcommand{\sla}[1]{/\!\!\!#1}

\jname{Annu. Rev. Nucl. Part. Sci.}
\jyear{2003}
\jvol{63}
\ARinfo{1056-8700/97/0610-00}

\title{Review of Top Quark Physics}

\markboth{Chakraborty, Konigsberg, Rainwater}
{Review of Top Quark Physics}
\author{
Dhiman ~Chakraborty, {\rm Northern Illinois University} \\
Jacobo ~Konigsberg, {\rm University of Florida} \\
David ~Rainwater, {\rm DESY}
}


\begin{abstract}
  We present an overview of Top Quark Physics - from what has been
learned so far at the Tevatron, to the searches that lie ahead at
present and future colliders.  We summarize the richness of the
measurements and discuss their possible impact on our understanding of
the Standard Model by pointing out their key elements and limitations.
When possible, we discuss how the top quark may provide a connection
to new or unexpected physics.
The literature on many of the topics we address is sizeable. We've
attempted to consolidate the most salient points and still give the
reader a complete, coherent overview. For more details the reader is
kindly referred to the corresponding seminal papers.

\end{abstract}

\maketitle

\section{Overview}
\label{sec:overview}

The discovery of the top quark at Fermilab's $p\bar{p}$ collider
Tevatron in 1995 by the CDF and D\O\ 
collaborations~\cite{discovery_d0_cdf} suggested the direct
experimental confirmation of the three-generation structure of the
Standard Model (SM) and opened up the new field of top quark physics.
Several properties of the top quark were studied at the Tevatron
during its first run.  These include measurement of $t\bar{t}$ pair
production cross section~\cite{exp_xsec} and kinematical
properties~\cite{kinematics_CDF,mt_D0_1lep,mt_D0_1lep_new,mt_D0_2lep},
top mass~\cite{mt_D0_CDF,mt_D0_1lep,mt_D0_1lep_new,mt_D0_2lep,mt_CDF},
tests of the SM via studies of $W$ helicity in top
decays~\cite{SM_CC_W_helicity_CDF} and spin correlations in $t\bar{t}$
production~\cite{Abbott:2000dt}, searches for electroweak production
of single top quarks~\cite{CDFsingletop,D0singletop} and for exotic
decays of top such as charged
Higgs~\cite{BSM_2HDM_H+_CDF,BSM_2HDM_H+_D0}, and flavor-changing
neutral currents~\cite{SM_FCNC_CDF}, etc. Precision of most of these
measurements are limited by statistical uncertainties because of the
small size of the data samples collected so far at the Tevatron (Run
1).  Run 2, currently underway, will increase the statistics by
approximately two orders of magnitude while the Large Hadron Collider
(LHC) and will be a true top factory, producing tens of millions of
top quarks every year (see Table~\ref{tab:collider_xs}).  The next
$e^+e^-$ Linear Collider (LC) would also have sufficient energy to
produce top quarks, and be ideally suited to precision studies of many
top quark properties.

\begin{figure}
\epsfxsize=300pt \epsfbox{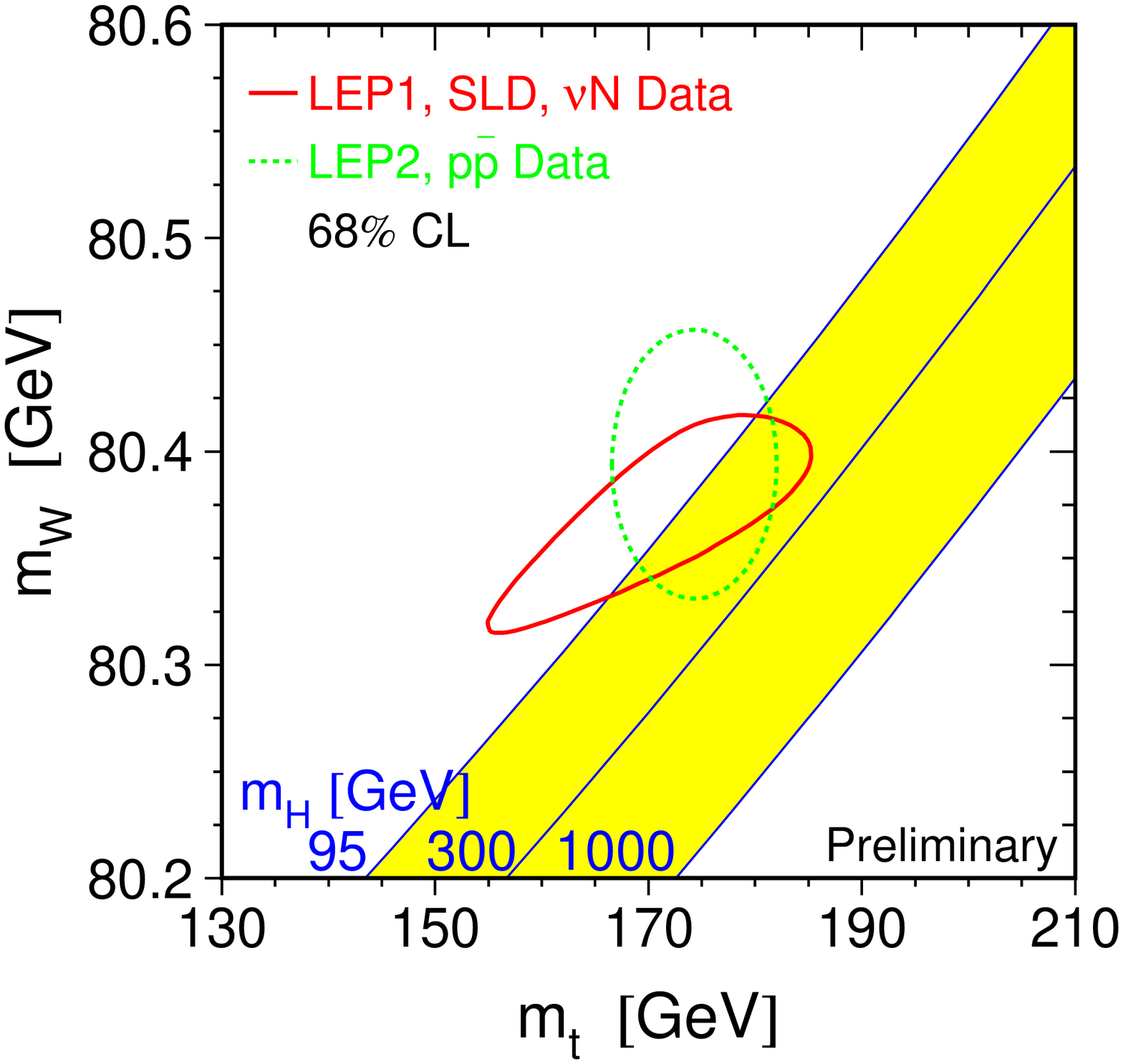}
\caption{The closed curves representing experimental measurements of 
  $M_W$ and $m_t$ constrain the SM Higgs mass. The shaded band shows
  the allowed combinations of $M_W$ and $m_t$ for different values of
  $M_H$.}
\label{fig:mw_mt_mh}
\end{figure}

The most striking observed feature that sets the top quark apart from
the other quarks is its very large mass.  Weighing in at $174.3\pm
5.1$ GeV~\cite{mt_D0_CDF}, it is about 35 times heavier than the next
heaviest quark, bottom ($b$), and is the heaviest elementary particle
known.  The top quark, $W$ and Higgs boson all contribute to radiative
terms in theoretical calculations of many observables that have been
measured with good precision by LEP, SLC and low-energy neutrino
scattering experiments.  Hence, precision measurement of $m_t$ and
$M_W$ constrain the mass of the SM Higgs boson, as shown in
Fig.~\ref{fig:mw_mt_mh}.

The vast swath of phase space available to the decay of such a heavy
quark gives it an extremely short lifetime, about $4\times 10^{-25}$~s
in the SM, ${\cal O}(10)$ times shorter than the characteristic
hadronization time of QCD, $\tau_{had} \approx 28\times 10^{-25}$~s.
As a result, the decay of top quarks offers a unique window on the
properties of a bare quark free from the long-range effects of QCD,
such as confinement.

The large mass of the top quark takes on even greater significance in
various extensions of the SM as particle spectra and flavor- or
mass-dependent couplings beyond the SM are contemplated: most such
particles are experimentally constrained to be heavier than all other
known fermions, but some may yet be lighter than the top quark and can
appear on-shell in its decays.  The top quark mass is also very close
to the energy scale of Electroweak Symmetry Breaking (EWSB).  Indeed,
its Yukawa coupling is curiously close to 1.  This raises the
possibility that perhaps there is more to it than its mass being
generated by the SM Higgs mechanism in the same way as postulated with
other fermions.

\subsection{Theoretical perspective}
\label{subsec:overview:top_SM_NP}

In the SM, the top quark is defined as the weak isospin partner of the
bottom quark.  As such, it is a spin-$\frac{1}{2}$ fermion of electric
charge $+\frac{2}{3}$ and transforms as a color anti-triplet under the
$SU(3)$ gauge group of strong interactions.  None of these quantum
numbers have yet been directly measured, although a large amount of
indirect evidence supports the SM assignments.  Precision measurements
of the $Z\to b\bar{b}$ partial width and forward-backward asymmetry at
LEP~\cite{PDG2002}, of $B^0$-$\bar{B}^0$ mixing, and limits on FCNC
decays of $B$ mesons require the existence of a particle with
$T_3=\frac{1}{2}$, $Q=\frac{2}{3}$ and mass near $170$ GeV, consistent
with the direct measurements by the Tevatron
experiments~\cite{B_meson_mix}.  The Tevatron $t\bar{t}$ production
cross section measurements are also consistent with theoretical
calculations for a particle with these attributes.  While Tevatron Run
2 will make more stringent tests, well enough to remove any doubt that
this is not the SM top quark, direct measurement of some of the top
quark quantum numbers will be possible only at the LHC and a LC.

The most pressing challenge in particle theory is to explain the
dynamics behind mass generation, which has two aspects: EWSB, whereby
the $W$ and $Z$ bosons acquire mass; and flavor symmetry breaking
(FSB), which splits the fermions into generations hierarchically
arranged in mass.  The SM accommodates both by postulating a
fundamental scalar field, the Higgs.  But this does not satisfactorily
explain the dynamics, and the Higgs sector runs into problems at high
energy scales.  One well-studied new physics explanation for this is
{\it technicolor} (TC), which postulates a new strong gauge
interaction at the TeV scale.  The top quark often plays a central
role in this class of models.  Another possibility is {\it
  supersymmetry} (SUSY), a new global space-time symmetry.  The
minimal SUSY model (MSSM) assigns a bosonic (fermionic) {\it
  superpartner} to every fermion (boson) in the SM, and predicts that
the lightest superfermion (sfermion) masses are close to that of their
SM partners.  The large top quark mass usually plays a central EWSB
role here as well.  Direct searches at LEP and Tevatron have set lower
limits on the masses of various SUSY particles~\cite{PDG2002}.  All of
these are well above $m_b$, but there is still enough room for SUSY
decays of the top quark.  A number of other theories postulate exotic
particles and interactions or new space-time dimensions for different
reasons, often cosmological.  In many of these, the large mass of the
top quark makes it a likely connection to new physics.

\subsection{The Experimental arena}

\subsubsection{Producing top}
\label{subsec:overview:colliders}

The only facilities where particles as massive as the top quark can be
produced at reasonable rates and studied effectively are symmetric
high-energy particle colliders, {\rm i.e.} where the center-of-mass
frame coincides with the laboratory frame.  To date, only the 1.8 TeV
incarnation of the Tevatron had sufficient energy to produce top
quarks.  The data collected during its Run 1 amounted to \mbox{$\sim
  600$} $t\bar{t}$ pair events in each of the detector experiments CDF
and D\O.  Only a small fraction of these passed the stringent
selection criteria imposed at the trigger level to suppress enormous
QCD backgrounds.  This was sufficient, however, to claim discovery of
top and make some initial measurements of its properties, principally
mass.  The current Run 2, with upgraded accelerator and detectors,
will result in perhaps a 100-fold increase in $t\bar{t}$ event yield
by 2008.  This will allow a more detailed examination of top,
sufficient to confirm its SM character, by drastically improving the
Run 1 measurements and making possible new ones.

Scheduled to start data collection in 2007, the 14 TeV $pp$ Large
Hadron Collider (LHC) is expected to deliver nearly eight million top
pair events to each of its two experiments, ATLAS and CMS, in the
first year alone.  The rate will increase by up to a factor of 10 in
subsequent years.  Even with a modest rate of acceptance, many rare
processes involving the top quark will become accessible.

Beyond the LHC, the most likely next collider would be a 500-1000~GeV
$e^+e^-$ linear collider. While the $t\bar{t}$ cross section would be
tiny compared to that at the LHC or even Tevatron, the integrated
luminosity would be large enough to produce at least half a million
top pair events in about five years of running. Moreover, there are
two main advantages to such a machine for precision studies. First,
$t\bar{t}$ production is an EW process. Theoretical calculations are
known to much higher precision in this case, and the absence of
enormous QCD backgrounds in experiment would yield extremely high
purity samples and nearly fully efficient event collection. Second,
because the center-of-mass energy of the colliding beams is exactly
known, top quarks could be reconstructed much more precisely. Variable
tuning of the beam energy would allow for production threshold scans,
giving access to super-precision measurements of mass and width.
Control over beam polarization would provide exceptionally detailed
couplings determinations. In short, a LC would be an ideal machine for
precision top quark physics. However, the main focus here is on recent
or approved experiments, i.e., the hadron colliders Tevatron and LHC.
We expect a future Annual Review article to concentrate on LC top
physics once such a facility is approved. Table~\ref{tab:collider_xs}
summarizes some key parameters for the colliders mentioned above.

\subsubsection{Detecting top}
\label{subsec:overview:detectors}

A top quark's production and decay vertices are separated by ${\cal
  O}(10^{-16})$~m, which exceeds the spatial resolution of any
detector by many orders of magnitude.  Detection of a top quark
therefore proceeds through identification and reconstruction of its
daughter particles.  Fortunately, the large top mass dictates it is
not produced highly relativistically.  Consequently, its much lighter
decay products have good angular separations and high momenta in the
laboratory frame. Most end up in the central region of the detector,
with $\vec{p}_T$, the momentum component perpendicular to the beamline
exceeding 20~GeV in magnitude~\footnote{Transverse momentum,
  $\vec{p}_T$, implies momentum measurement with a magnetized tracker
  (e.g. for electrons and muons) while transverse energy, $\vec{E}_T$,
  implies calorimeter energy measurement (e.g. for jets). The two have
  the same physical interpretation, but different resolutions.}.

\begin{figure}
\epsfxsize=300pt \epsfbox{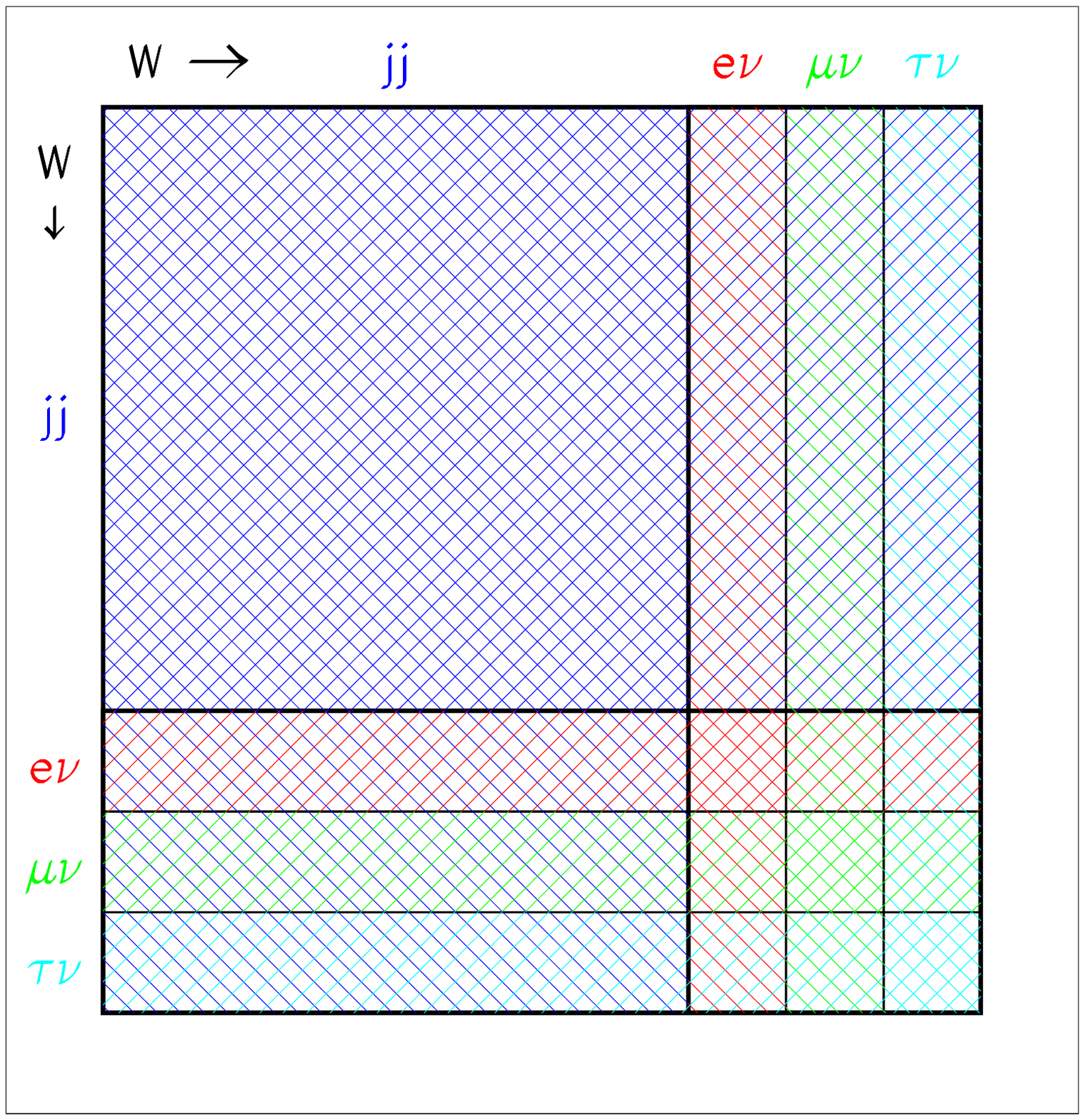}
\caption{Branching fractions of $t\bar{t}$ due to the various 
  subsequent $W$ decays.  All final states have an additional
  $b\bar{b}$ pair from the top decays.}
\label{fig:pie}
\end{figure}

Top decay products span the entire spectrum of quarks and leptons.
Within the SM, the top quark decays almost exclusively into $Wb$.  The
$W$ decays almost instantaneously (lifetime $\sim 3\times 10^{-25}$~s)
either {\it leptonically} into a lepton-neutrino pair: $B(W\rightarrow
\ell \bar\nu_\ell)=\frac{1}{3}$, ($\ell= e,\mu,\tau$ with equal
probabilities) or {\it hadronically} into a quark-antiquark pair:
$B(W\rightarrow q_1\bar q_2)=\frac{2}{3}$, ($q_1(q_2)=u(d), c(s)$ with
equal probabilities).  Hadronic final states manifest themselves as a
shower of particles, called a {\it jet}.  If the $W$ decays
leptonically, then the charged lepton can be identified with relative
ease, $\tau$ being an exception, while neutrinos escape direct
detection.  A graphical representation of the various SM branching
fractions of top pairs is shown in Fig.~\ref{fig:pie}.  Normally in
the experimental context of hadron colliders, only $e,\mu$ are
referred to as {\it leptons}, since $\tau$ final states behave so
differently.

This large and complex set of final state permutations has significant
implications for data collection.  Although a multilayered hardware
and software triggering system is carefully designed to retain as many
of the most interesting events as possible, and the detector is almost
hermetic, some fraction of top events will be lost depending on the
decay mode and distribution, as well as the priorities of the
experimental program. A brief account of the major issues for
particles entering the detector is in order:
\begin{itemize}
  
\item Electrons are recognized with about $90\%$ efficiency by their
  short interaction length leading to a compact shower in the
  calorimeter and an associated track of matching momentum in the
  central tracking volume of the detector.
  
\item Muons are highly penetrating particles that are distinguished by
  their minimum-ionizing trail all the way through being the only
  particles to reach the outermost detector layers, with about $90\%$
  efficiency.
  
\item Neutrinos escape direct detection because of their tiny weak
  interaction.  Since the beam-axis component of net event momentum
  varies over a wide range at a hadron collider, only the transverse
  component of invisible particles' total momenta, \vmpT (\vmET), can
  be inferred in any given event.  Simplistically, it is the negative
  vector sum of observed particles' transverse momenta. The \vmET
  resolution depends strongly on the content and topology of an event.
  
\item Detection of $b$ quarks is particularly important in selection
  of top event candidates since most QCD events don't contain them, so
  their identification reduces backgrounds considerably.  A $b$
  immediately hadronizes, but typically travels about half a
  millimeter from the primary interaction vertex before decaying into
  a jet containing multiple charged particles.  Such a displaced decay
  vertex can be isolated using a good vertex detector by extrapolating
  the tracks associated with the jet to a common origin ({\it
    secondary vertex tagging}).  Jets initiated by gluons and lighter
  quarks (except sometimes $c$) are rarely associated with a secondary
  vertex.  Additionally, about $20\%$ of the time a $b$ jet contains a
  lepton which typically has a lower momentum than a prompt lepton
  from a $W$ decay.  This offers an alternative means for tagging a
  $b$ quark jet ({\it soft lepton tagging}). Overall, $b$ quarks can
  be identified about $60\%$ of the time.
  
\item Tau leptons decay leptonically $36\%$ of the time and
  hadronically $64\%$.  The leptonic decays result (in addition to two
  neutrinos) in an electron or a muon that are typically softer than
  those from $W$ decays.  Apart from a very small impact parameter
  that is difficult to measure, $W\to \tau\bar\nu_\tau \to
  \ell\bar\nu_\ell\nu_\tau\bar\nu_\tau$ ($\ell=e,\mu$) decays cannot
  really be singled out from $W\to \ell\bar\nu_\ell$ in top events,
  and are automatically accounted for in the measurements with
  electron and muon final states.  The hadronic modes need special
  consideration: $\sim 76\%$ of these yield a single charged daughter
  ({\it 1-prong}) and $\sim 24\%$ yield 3 ({\it 3-prong}). Good
  pattern-recognition algorithms can exploit the low charge
  multiplicity and characteristic features of the associated narrow
  shower in the calorimeter to separate hadronic $\tau$ decays from
  the copious QCD background.  The associated neutrino carries away a
  significant fraction of the $\tau$ momentum, making its estimation
  dependent on the distribution of other objects in the event.
  Overall, the identification efficiency of hadronic tau decays is
  about $50\%$.
  
\item Jets initiated by gluons and lighter quarks have nearly full
  detection efficiency, although establishing their partonic identity
  on an event-by-event basis is not possible as they hadronize into
  overlapping states.  Subtle differences in profiles of gluon and
  quark jets my be discernible on a statistical basis. If so, it would
  be very useful to top quark studies since all jets from top decays
  are quark initiated (discounting final-state radiation), while jets
  in the QCD background are predominantly gluon-initiated. This
  possibility requires further studies in the context of hadron
  colliders.  Jets arising from gluons and lighter quarks will be
  misidentified as a $b$($\tau$) at a rate of only about 1/200.  They
  fake an electron or muon even more rarely, at about the 1/2000
  level.
  
\end{itemize}

Top quark decays are no less varied in scenarios beyond the SM.
Therefore, identification of all of these {\it objects} as well as
accurate and precise measurement of their momenta are key to studies
of the top quark.  Detailed description of the detector design and
performance specifications are available
elsewhere\cite{Tevatron_detectors_Run1,Tevatron_detectors_Run2,LHC_detectors}.


Detailed comparisons of the experimental measurements of the nature of
top quark production (cross section and kinematics), decay (partial
widths, angular correlations among decay products, and so on), and
other properties (mass, discrete quantum numbers, etc.), with those
theoretically predicted are important probes for new physics.  It is a
challenge for theorists and experimentalists alike to perform
calculations and measurements at the highest possible level of
precision. For readers interested in greater detail, especially from
an experimenter's point of view, we strongly recommend two excellent
articles: Refs.~\cite{Top_at_tevatron} for top quark physics at the
Tevatron, and Ref.~\cite{Beneke:2000hk} for that at the LHC.  Earlier
accounts of the discovery of the top quark can be found in
Refs.~\cite{Wimpenny:1996dz,Campagnari:1996ai}.

\begin{table}[htb]
\def~{\hphantom{0}}
\caption{Operation parameters of present and future colliders, and 
cross sections for some important processes. For $\sigma({t\bar{t}})$,
(a) is the complete NLO+NLL calculation, while (b) is the partial 
NNLO+NNLL calculation, discussed in Sec. 2.1.1. The integrated 
luminosities are per experiment.} 
\label{tab:collider_xs}
\vspace{3mm}
\begin{tabular}{|c|c|c|c|c|}
\toprule
Collider & Tevatron Run 1&Tevatron Run 2& LHC & LC \\
\colrule \hline
type & $p\bar{p}$ & $p\bar{p}$& $pp$ & $e^+e^-$ \\
Run period& 1992-1996 & 2001-2008(?) & 2007-? & 2015(?)-? \\
\colrule \hline
$E_{\rm CM}$ (TeV) & 1.80 & 1.96 & 14.0 & $<2m_t$ - $\sim$1.0 \\
$\mean{\cal L}$ (cm$^{-2}$s$^{-1}$) & $1\times 10^{31}$ & $1\times 10^{32}$ & $10^{33}$ - $10^{34}$ & $2\times 10^{34}$ \\
$\int{\cal L}dt$ (fb$^{-1}$)
& 0.125 & 6.5 - 11 & $\sim$300 & $\sim$1000 \\
\hline
$\sigma_{\rm total}$ (pb) &
$\sim 10^{11}$ & $\sim 10^{11}$ & $\sim 10^{11}$ & 
${\cal O}(10)$ \\
$\sigma(b\bar{b})$ (pb) & 
$\sim 2\cdot 10^7$ & $\sim 3\cdot 10^7$ & 
$\sim 3\cdot 10^8$ & ${\cal O}(1)$ \\
$\sigma(WX)$ (pb) & 
$\sim 3\cdot 10^{4}$ & $\sim 4\cdot 10^{4}$ & 
$\sim 2\cdot 10^{5}$ & ${\cal O}(1)$ \\
$\sigma(t\bar{t})(a)$ (pb) & 
$5.06^{+0.13}_{-0.36}$ & $6.97^{+0.15}_{-0.47}$ & $825^{+58}_{-43}$ & $\sim0.8$ \\
$\sigma(t\bar{t})(b)$ (pb) & 
$5.8\pm 0.4$ & $8.0\pm 0.6$ & - & - \\
$\sigma$(single $t$) (pb) & 
$1.08\pm 0.01$ & $1.50\pm 0.02$ & $315^{+8}_{-2}$ & $\sim 0$ \\
\botrule
\end{tabular}
\label{table:collider_params}
\end{table}

\section{Top Quark Production}
\label{sec:production}

At hadron colliders two distinct SM production mechanisms are
possible: dominant $t\bar{t}$ pair production, via the strong
interaction; and single-top production via the electroweak (EW)
interaction. As we shall see, detailed comparison between experimental
measurements of physical observables related to top quark production,
and SM predictions, is an important probe for new physics.

\subsection{Pair production}
\label{subsec:production:SM_QCD}

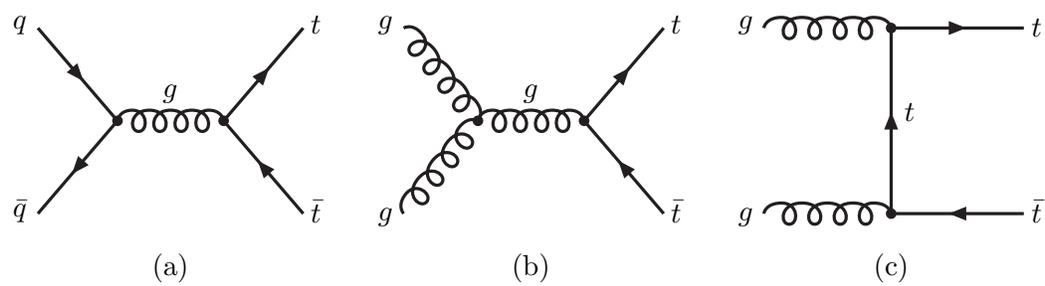
\begin{figure}
\begin{center}
        \begin{picture}(90,85)(0,0)
        \SetOffset(50,40)
        \SetWidth{1.2}
        \ArrowLine(-20,0)(-50,-35)
        \ArrowLine(-50,35)(-20,0)
	\Vertex(-20,0){2}
        \Gluon(-20,0)(20,0){4}{4}
	\Vertex(20,0){2}
        \ArrowLine(50,-35)(20,0)
        \ArrowLine(20,0)(50,35)
        \Text(-57,33)[b]{$q$}
        \Text(-57,-30)[t]{$\bar{q}$}
        \Text(0,7)[b]{$g$}
        \Text(55,33)[b]{$t$}
        \Text(55,-30)[t]{$\bar{t}$}
        \Text(0,-50)[t]{(a)}
        \end{picture}
\hskip 1.5 cm
        \begin{picture}(90,80)(0,0)
        \SetOffset(50,40)
        \SetWidth{1.2}
        \Gluon(-48,-35)(-20,0){4}{4}
        \Gluon(-48,35)(-20,0){4}{4}
	\Vertex(-20,0){2}
        \Gluon(-20,0)(20,0){4}{4}
	\Vertex(20,0){2}
        \ArrowLine(50,-35)(20,0)
        \ArrowLine(20,0)(50,35)
        \Text(-55,33)[b]{$g$}
        \Text(-55,-33)[t]{$g$}
        \Text(0,7)[b]{$g$}
        \Text(55,33)[b]{$t$}
        \Text(55,-30)[t]{$\bar{t}$}
        \Text(0,-50)[t]{(b)}
        \end{picture}
\hskip 1.5 cm
        \begin{picture}(90,80)(0,0)
        \SetOffset(50,40)
        \SetWidth{1.2}
        \Gluon(-48,-35)(0,-35){4}{4}
        \Gluon(-48,35)(0,35){4}{4}
	\Vertex(0,-35){2}
	\Vertex(0,35){2}
        \ArrowLine(0,-35)(0,35)
        \ArrowLine(50,-35)(0,-35)
        \ArrowLine(0,35)(50,35)
        \Text(-55,33)[b]{$g$}
        \Text(-55,-33)[t]{$g$}
        \Text(7,0)[b]{$t$}
        \Text(55,32)[b]{$t$}
        \Text(55,-30)[t]{$\bar{t}$}
        \Text(0,-50)[t]{(c)}
        \end{picture}

\end{center} 
\caption{Leading order Feynman diagrams for $t\bar{t}$ production via
  the strong interaction.}
\label{fig:tt_prod}
\end{figure}

In the SM, $t\bar{t}$ pairs are produced via quark-antiquark
($q\bar{q}$) anihilation and gluon fusion. Figure~\ref{fig:tt_prod}
shows the corresponding leading order (LO) Feynman diagrams.

The total tree level (Born approximation) $t\bar{t}$ cross section at
hadron colliders is a convolution of the parton distribution functions
(PDFs) for the incoming (anti)protons and the cross section for the
partonic processes $q\bar{q},gg \to t\bar{t}$:
\begin{equation}
\sigma(s,m_t^2) = 
\sum_{i,j} \int_0^1 dx_1 \int_0^1 dx_2 \,
f_i(x_i,\mu^2_f) \, f_j(x_j,\mu^2_f) \;
\hat\sigma_{ij}(\hat{s},m_t,\alpha_s(\mu^2_r)),
\label{eq:tt_xsec}
\end{equation}
where $i,j$ are the possible combinations of incoming gluon or
quark-antiquark pairs and $f(x,\mu^2_f)$ are the PDFs, evaluated at
some factorization scale $\mu_f$ corresponding to a scale in the
problem, such as $m_t$, and a value $x$ which is the fraction of
incoming (anti)proton energy that the parton carries.  The partonic
subprocess cross sections, integrated over phase space, are functions
of the center-of-mass energy $\sqrt{\hat{s}}$, the top quark mass
$m_t$, and the QCD strong coupling constant $\alpha_s$ evaluated at a
renormalization scale $\mu_r$, also typically taken to be one relevant
to the process, e.g. $m_t$, but it does not have to be the same as
$\mu_f$. At higher orders, the partonic cross section also depends on
$\mu_f,\mu_r$:
$\hat\sigma_{ij}(\hat{s},m_t,\mu_f,\mu_r,\alpha_s(\mu^2_r))$.

At the Tevatron, $t\bar{t}$ production occurs close to, but not quite
at threshold. The maximum of $d\sigma_{t\bar{t}}/d\hat{s}$ occurs
around 3/2 the threshold value, and the average speed of the top
quarks is about $\beta\approx 0.5$. If for threshold we set
$x_i\approx x_j = x_{\rm thr}$, from $\hat{s} = x_i x_j s$ we obtain
$x_{\rm thr}\approx{2m_t\over\sqrt{s}}$.  In Tevatron Run~1, $x_{\rm
  thr} \approx 0.2$, where the quark distribution functions are
considerably larger than that for the gluon, $q\bar{q}(gg)\to
t\bar{t}$ accounted for $90\%(10\%$) of the cross
section~\footnote{For the partonic cross sections, $\sigma_{gg} >
  \sigma_{q\bar{q}}$, but parton densities are the dominant effect.}.
In Run~2, $\sqrt{s} = 2.0$~TeV, the total cross section is about
$40\%$ larger, with $85\%(15\%)$ coming from an initial $q\bar{q}(gg)$
pair.  At the LHC the situation is reversed: $x_{\rm thr}\sim 0.025$,
a regime where gluons dominate, so the $q\bar{q}(gg)$ contributions
are about $10\%(90\%)$.  Table~\ref{tab:collider_xs} summarizes the
$t\bar{t}$ cross sections at the Tevatron, LHC and a LC, compared to
other important SM processes.  At the Tevatron, roughly one in
$10^{10}$ collisions produces top quark pairs.  In Run~1 the average
production rate was $\sim 5\cdot 10^{-5}$~Hz, expected to reach $\sim
7\cdot 10^{-4}$~Hz in Run~2.  In comparison, the rate will be about
10~Hz at the LHC, a true ``top factory".

The uncertainty in $\sigma^{LO}_{t\bar{t}}$ at hadron colliders is
large, $\sim 50\%$.  The primary source centers around the scale
choices $\mu_f$ and $\mu_r$, and their effects on $\alpha_s$.
Furthermore, $\alpha_s$ is relatively large, so additional terms in
the perturbative expansion for the cross section can be significant.
These issues can be addressed by calculating the cross section at
next-to-leading order (NLO) in perturbation theory, which we discuss
in the next section.  Additional, smaller sources of uncertainty are
the PDFs and the precise values of $m_t$ and $\alpha_s(M^2_Z)$.  At
the Tevatron the cross section sensitivity due to PDFs is small mainly
because the process is driven by the well measured quark
distributions.  This is not the case at the LHC, where a $\sim 10\%$
uncertainty in $\sigma_{t\bar{t}}$ comes from the PDF for the dominant
$gg$ component.

\subsubsection{Higher order corrections and theoretical uncertainties}
\label{sec:NLO}

At LO the $t\bar{t}$ cross section is usually evaluated for
$\mu_f=\mu_r = m_t$, as $m_t$ is the only relevant scale in the
problem (one could also argue for $2m_t$ for $\alpha_s$, but $\mu_r =
\mu_f$ is the more common choice).  Since this is much larger than the
scale of QCD confinement, $\Lambda_{QCD}\sim 200~{\rm MeV}$, the
calculation can be trusted to behave perturbatively.  But what does
the scale choice signify?  After all, both PDFs and $\alpha_s(M^2_Z)$
are data extracted from experimentally measured cross sections.
However, they are based on processes very different from those we wish
to consider at hadron colliders.  We have to let $\alpha_s$ {\it run}
and the PDFs {\it evolve} from the scales relevant in extraction to
the scales relevant for application.  The calculation of the process
under consideration is separated into two parts: the perturbative hard
scattering (here, $q\bar{q},gg\to t\bar{t}$), and the perturbatively
resummed PDF evolution which uses non-perturbative input.  To this
end, the scales $\mu_r,\mu_f$ are introduced to separate the
perturbative and non-perturbative parts of the calculation.

By construction, physical observables in a renormalizeable field
theory do not depend on a scale.  But this is true only to all orders
in perturbation theory, which is impossible to calculate. At fixed
order, the scale independence is not realized.  Higher orders help
restore this, removing bit by bit the scale dependence we artificially
introduced.  Varying the scale at a given order gives one an idea of
the residual calculational uncertainty.

In a higher-order calculation, all diagrams that contain the same
order in the relevant coupling must be included. Here, this is
$\alpha_s$.  Thus, the full ${\cal O}(\alpha_s^3$) NLO
calculation~\cite{tt_NLO} includes both real parton emission and
virtual (loop diagram) corrections, even though the different parts do
not contain the same number or even type of final state particles.
The NLO corrections increase $\sigma_{t\bar{t}}$ by about $30\%$, with
the uncertainty from varying the scale choice reduced to about $12\%$.

An important point to note is that the order of the hard scattering
process evaluated must match that of the PDF set used. At each higher
order in $\alpha_s$, there are strong cancellations between terms in
the PDF evolution and in the hard scattering real emission, which come
from the artificial dependence on $\mu_f$ introduced by factorizing
the problem in the first place. For NLO calculations, NLO PDFs must be
used; for LO calculations such as parton shower Monte Carlo (MC), LO
PDFs must be used.  Noncompliance can introduce large errors.

The NLO calculation of $\sigma_{t\bar{t}}$ experiences large
logarithms $\sim\alpha_s\log^2{\beta}$, where $\beta$ is some
definition of the threshold dependence (which can vary at NLO),
arising from real emission of a soft gluon.  As $\beta\to 0$ at
threshold, the calculation becomes unstable.  Fortunately, real
radiation there is restricted by phase space, so soft gluons
approximately {\it exponentiate}: an $(\alpha_s\log^2{\beta})^n$ term
appears at all orders in perturbation theory, with a coefficient at
each order of ${1\over n!}$ from permutations over identical gluons,
resulting in a series that is simply an exponential containing
$\alpha_s\log{\beta}$.  Calculating it is called {\it resumming} the
large logs.  This behavior is a direct consequence of soft gluon
emission in QCD factorizing both in the matrix element and in phase
space.  A leading-log (LL) resummation takes care of the
$(\alpha_s\log^2{\beta})^n$ series, a next-to-leading-log (NLL)
resummation the $(\alpha_s(\alpha_s\log^2{\beta}))^n$ series, and so
on.  This is an overly simplistic picture, but gives one an idea of
what resummation calculations address.

According to one recent NLO+NLL complete resummation
calculation~\cite{Bonciani:1998vc}, with PDF-updated results for the
LHC in Sec.~2 of Ref.~\cite{Beneke:2000hk}, resummation effects are at
the ${\cal O}(5\%)$ level for both the Tevatron and LHC.  Results are
$\sigma_{t\bar{t}}= 5.06(6.97)$ pb for $p\bar{p}$ collisions at
$\sqrt{s}=1.8(2.0)$ TeV and 825 pb for $pp$ collisions at 14 TeV,
where the uncertainties are from scale variation. Another $\sim 6\%$
contribution comes from PDFs and $\alpha_s$.

Another recent Tevatron-only study~\cite{Kidonakis:2001nj} is a
partial NNLO+NNLL calculation, where they expand the exponential
expression to the first three powers of the large logs at ${\cal
  O}(\alpha_s)$ and ${\cal O}(\alpha^2_s)$. This study finds a
$5-20\%$ uncertainty depending on the $t\bar{t}$ kinematics
considered, and averages the results to construct total estimates of
$\sigma_{t\bar{t}}({\rm 1.8 TeV}) = 5.8\pm 0.4\pm 0.1$~pb and
$\sigma_{t\bar{t}}({\rm 2.0 TeV}) = 8.0\pm 0.6\pm 0.1$~pb, where the
first uncertainty is due to kinematics and the second is from scale
uncertainty.

The Tevatron results of Refs.~\cite{Bonciani:1998vc,Kidonakis:2001nj}
are not necessarily contradictory, since they use different methods
that selectively incorporate different higher-order terms.  For
uncertainties at the LHC, the relation is~\cite{Beneke:2000hk}
${\delta\sigma\over\sigma}\sim 5{\delta m_t \over m_t}$, i.e., if
1~GeV in $\delta m_t$ is achievable, then the cross section should be
known to about $3\%$ experimentally.  This makes improvements in
$\sigma^{NLO}_{t\bar{t}}$ desireable, although a complete NNLO
calculation is not likely to be completed soon.  At the very least, it
would be useful to have an improved understanding of PDFs, such as a
more sophisticated PDF-uncertainty analysis.

Besides the soft gluon effects, Coulomb effects may enhance or deplete
the cross section near threshold. However, these are found to be
negligibly small for $t\bar{t}$ production at both the Tevatron and
LHC~\cite{Catani:1996dj}, much smaller than the inherent uncertainty
in the NLO+NLL calculations. The same holds true for EW corrections,
found to be $-0.97\%$ to $-1.74\%$ of $\sigma^{LO}_{t\bar{t}}$ for
$60<M_H<1000$~GeV~\cite{Beenakker:1993yr}.

\subsubsection{Experimental measurements: cross-sections, kinematics}

We now turn to the question of how to measure experimentally the
$t\bar{t}$ production cross section and how accurate these
measurements are expected to be.

\begin{figure}
\begin{center}
	\begin{picture}(200,190)(0,0)
        \SetOffset(70,100)
        \SetWidth{1.2}
        \ArrowLine(-25,0)(-70,-35)
        \ArrowLine(-70,35)(-25,0)
	\Vertex(-25,0){2}
        \Gluon(-25,0)(25,0){4}{5}
	\Vertex(25,0){2}
        \ArrowLine(70,-25)(25,0)
        \ArrowLine(25,0)(70,25)
        \Text(-75,31)[b]{$q$}
        \Text(-75,-29)[t]{$\bar{q}$}
        \Text(0,7)[b]{$g$}
        \Text(50,18)[b]{$t$}
        \Text(50,-18)[t]{$\bar{t}$}
        \SetOffset(140,75)
        \ArrowLine(70,-60)(0,0)
        \Text(75,-65)[b]{$\bar{b}$}
        \Photon(0,0)(40,0){3}{4}
        \Text(20,7)[b]{$W^-$}
        \ArrowLine(40,0)(70,15)
        \Text(75,12)[b]{$e$}
        \ArrowLine(70,-15)(40,0)
        \Text(78,-20)[b]{$\bar\nu_e$}
        \SetOffset(140,125)
        \ArrowLine(0,0)(70,60)
        \Text(75,66)[t]{$b$}
        \Photon(0,0)(40,0){3}{4}
        \Text(20,-7)[t]{$W^+$}
        \ArrowLine(40,0)(70,15)
        \Text(75,19)[t]{$u$}
        \ArrowLine(70,-15)(40,0)
        \Text(75,-10)[t]{$\bar{d}$}
	\end{picture}

\end{center}
\caption{Leading order Feynman diagram of single lepton decay of a 
  $t\bar{t}$ event.}
\label{fig:ttdecay}
\end{figure}
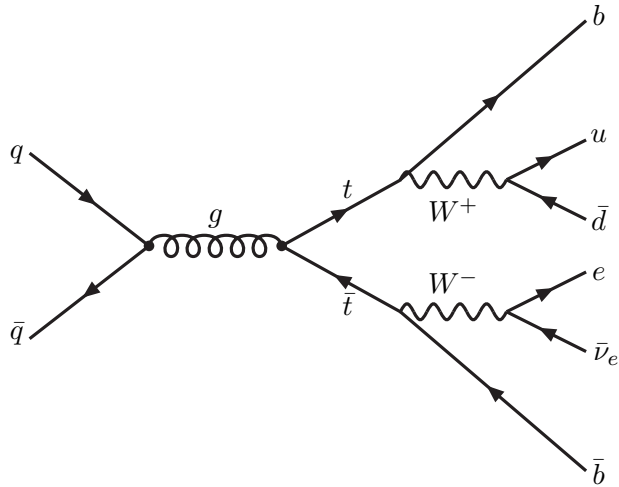

Within the SM, the top quark decays almost exclusively into a $W$
boson and a $b$ quark.  The channels and branching fractions for
$t\bar{t}$ decays can be readily derived from those for $W$ decays
given in Sec.~\ref{subsec:overview:detectors}.  Because of the
uniqueness of their experimental detection, channels involving $\tau$
leptons are usually treated separately.  In the context of object
identification in the detector, unless noted otherwise, a ``lepton''
normally refers to $e$ or $\mu$.  Thus, the $t\bar{t}$ final state is
categorized as ``dilepton'' (branching fraction = $5\%$),
``single-lepton (plus jets)'' ($30\%$) and ``all-hadronic'' ($44\%$)
depending on whether both, only one, or neither of the two $W$ bosons
decay leptonically into an electron or a muon and the corresponding
neutrino (Fig.~\ref{fig:ttdecay}).  The remaining $21\%$ involves
$\tau$ leptons: $6\%$ for ``$\tau$-dilepton'' ($e\tau$, $\mu\tau$,
$\tau\tau$) and $15\%$ for $\tau$+jets.

\begin{itemize}
\item {\bf Modeling $t\bar{t}$ production}
\end{itemize}

Accurate simulation of collision events is critical to the
understanding of how to derive reliable physics measurements from the
detector data.  Experimentalists use MC generators such as {\sc
  pythia}~\cite{PYTHIA}, {\sc herwig}~\cite{HERWIG} or {\sc
  isajet}~\cite{ISAJET} to model $t\bar{t}$ production in hadron
collisions.  These include approximate treatments of higher order
perturbative effects (initial and final state gluon radiation),
hadronization of the final state partons, underlying event, and
secondary particle decays.  They begin by using an exact matrix
element calculation (QCD or EW) of the hard scattering process, such
as $q\bar{q}\to t\bar{t}$, then simulate the emission of additional
partons from the incoming and outgoing partons in the hard process.
This is done with a {\it parton shower} algorithm evolving the emitted
parton energies downwards to a cutoff point, where hadronization takes
over.

A more detailed description of these MC programs can be found e.g. in
Ref.~\cite{Beneke:2000hk}. The events these generators produce are
then combined with the simulation of the detectors' response to the
final state particles. Event selection {\it cuts} can then be studied
to understand how best to optimize the signal acceptance while
reducing backgrounds from other physics processes that can fake a
$t\bar{t}$ signature.

There are ``small" discrepancies between some of the predictions in
these MC programs.  For example, {\sc pythia} and {\sc herwig} differ
in the amount of gluon radiation that they
introduce~\cite{Mrenna97,Orr97}.  Tests comparing distributions from
the MC predictions to the NLO calculations can be found e.g. in
Ref.~\cite{Frixione:1995fj} which concludes that in the low-$p_T$
region {\sc herwig} more closely approximates the NLO calculations.

It is clear that as larger $t\bar{t}$ datasets are gathered by the
experiments, more detailed comparisons between data and MC predictions
will be feasible and a positive feedback loop will be established.
This will lead to improved understanding of mechanisms behind the more
subtle aspects of $t\bar{t}$ production.  Accurate modeling will be
critical in detecting any possible deviation from the SM predictions.

\begin{itemize}
\item {\bf Event selection and backgrounds}
\end{itemize}

It is important to understand how the rare $t\bar{t}$ events are
selected from the flood of other events generated in hadron
collisions, and how they are separated from backgrounds that pass the
same selection criteria.  We discuss the experiments at the Tevatron
and then point out the differences, if any, for the LHC.

As would be expected in the decay of a massive, slow-moving particle
($\beta \ll 1$) into almost massless ones, the final state particles
in top decay typically carry large transverse momentum in the lab
frame ($p_T> 15-20$~GeV), and often go into the more central part of
the detector ($|\eta|<\sim 2.5$)~\footnote{$\eta = {1\over
    2}{E+p_z\over E-p_z}$ is called the {\it pseudorapidity}, which
  for massless particles is $\eta =
  -\ln\left(\tan{\displaystyle\frac{\theta}{2}}\right)$.}  Therefore,
regardless of channel, the first experimental criteria for detecting
top events is requiring high $p_T$ for all decay products 
This requirement goes a long way in suppressing
backgrounds, especially processes with jets from QCD radiation, which
have an exponentially falling $E_T$ spectrum, and processes in which
\mET is an artifact of instrumental imprecision, not the escape of
real, high-$p_T$ neutrinos.  Other topological cuts, such as requiring
that the leptons and \mET are isolated from jet activity and more
global event variables such as {\it scalar $E_T$} ($H_T$, the scalar
sum of $E_T$ of all observed objects), sphericity and
aplanarity~\footnote{Defined e.g. in Ref.~\cite{Barger:nn}.} help
enhance the signal-to-background ratio ($S:B$).  The latter two are
variables calculated from the eigenvalues of the normalized momentum
tensor.  Aplanarity ${\cal A}$, proportional to the smallest of the 3
eigenvalues, measures the relative activity perpendicular to the plane
of maximum activity.  Sphericity ${\cal S}$, proportional to the sum
of the two smaller eigenvalues, measures the relative activity in the
plane of minimum activity.  Top quark events typically have larger
values of $H_T$, ${\cal S}$, ${\cal A}$.  Finally, the $b$-tagging
requirement eliminates most non-top QCD contamination of the signal,
about a 100-fold reduction, compared to $\sim 75\%$ of the top events
yielding at least one tagged $b$-jets~\footnote{The efficiencies of
  Sec.~\ref{subsec:overview:detectors} have to be moderated by the
  fiducial acceptance of the detector.}.  Tagging heavy-flavor jets
with soft leptons helps disentangle systematic uncertainties of the
QCD heavy-flavor content.

Remaining backgrounds in the all-hadronic channel arise mainly from
QCD multi-jet production, in which $b$ tags from real heavy-flavor
quarks (mostly $b$, but also some $c$) or from fakes (gluons or light
quarks) are present.  The $S:B$ ranges from 1:5 to 1:1 depending on
details of the selection.  In the single-lepton channel the most
copious background is from $W$+jets events, before $b$-tagging, and
from $W$+heavy-flavor after.  The $S:B$ after $b$-tagging is typically
between 1:1 and 4:1, again depending on the exact criteria.  For
dileptons, $S:B\approx 1:2$, even without $b$-tagging, with
backgrounds coming mainly from $WW$, $Z\to\tau^+\tau^-$ and Drell-Yan
production, all with additional jets from QCD radiation.  The
background in this case becomes negligible if the requirement of
$b$-tagging is added.  This is because these backgrounds are all
either EW suppressed, or arise only from several small branching
fractions successively.  Including branching fractions and
efficiencies of the full chain of selection criteria, only a few
percent of the $t\bar{t}$ events produced in the collisions make it to
the final sample.  In Run 1 an estimated $5\%$ made it to the
all-hadronic candidate pool, about $5\%$ to the single-lepton, and
only about $1\%$ to the dilepton pool.

An excess of about 10 dilepton events over an expected background of 4
events was observed in the combined data samples of CDF and D\O.  Some
of these candidates have been suggested as having unusual
kinematics~\cite{Barnett:1996zr}; Run 2 should resolve this question.
In the single-lepton channel, with [without] b-tags, an excess of
about 60 [10] events was observed over an expected background of about
40 [9].  In the all-jet channel D\O\ [CDF] observed an excess of 16
[43] events over a background of about 25 [144].

At the LHC, very pure signals should be obtained in the dilepton and
single-lepton channels.  For 10 fb$^{-1}$, with similar selection
criteria as those used at the Tevatron, about 60,000 $b$-tagged
dilepton events are expected, with a $S:B\approx
50$~\cite{Beneke:2000hk}.  In the single-lepton channel, this will be
close to one million $b$-tagged events.  Since the cross section for
QCD $W$+jets grows more slowly with collision energy than does
$t\bar{t}$, $S:B\approx 20$ should be possible.  However, extracting
such a clean signal on the all-jets channel out of overwhelming QCD
background is not deemed feasible.  Ongoing studies selecting on more
sophisticated kinematical variables and using multivariate
discriminants show a paltry $S:B\approx 1:6$.

\begin{figure}
  \epsfxsize=400pt \epsfbox{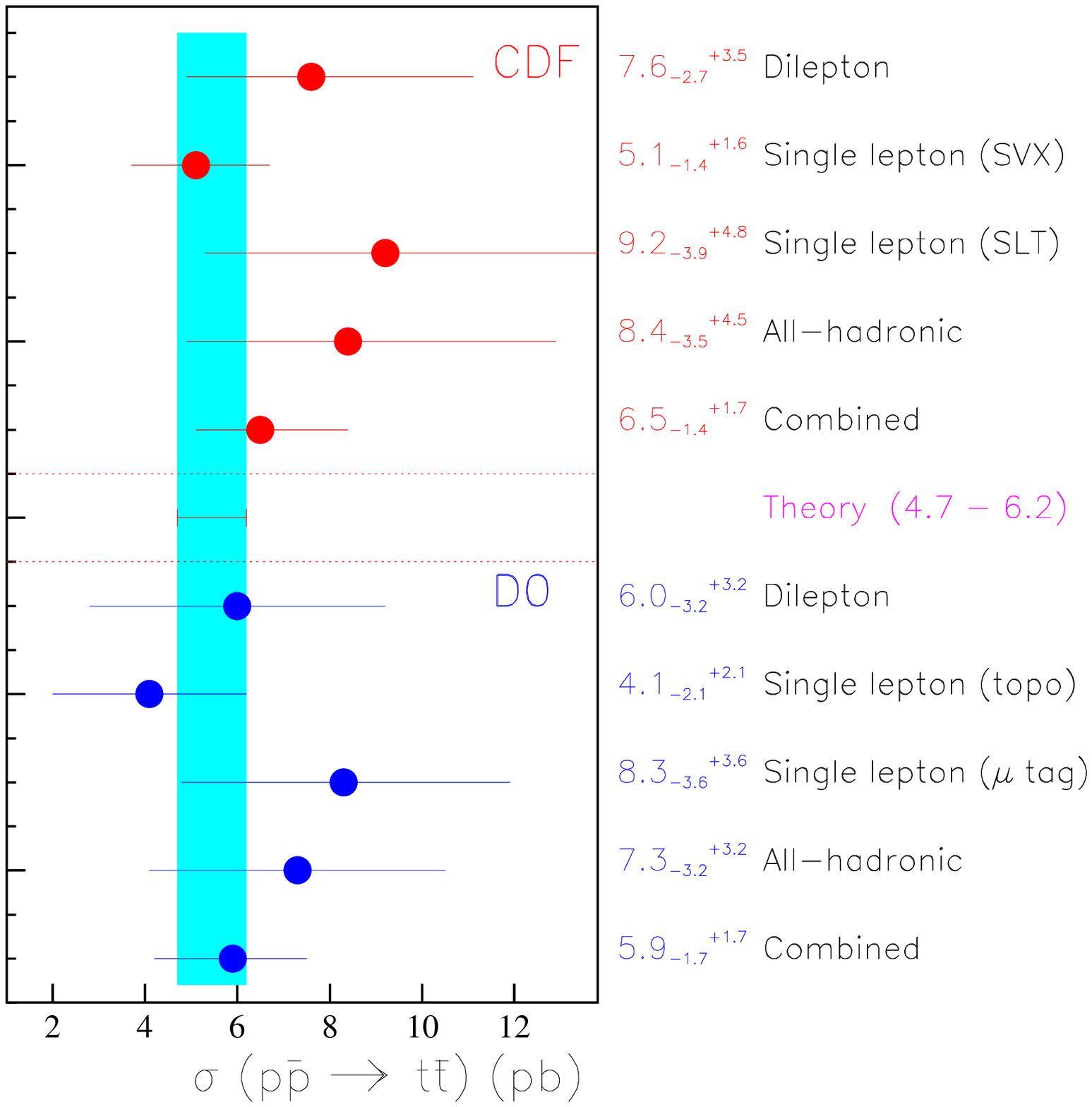}
\caption{CDF and D\O\ cross section results for $t\bar{t}$ production
  at the Fermilab Tevatron, Run 1, overlaid with the theory
  prediction.  For the latter, we take the entire band covered by
  both the NLO+NLL and partial NNLO+NNLL predictions (see text).}
\label{fig:xsec_results}
\end{figure}

Figure~\ref{fig:xsec_results} shows the $t\bar{t}$ cross section
results individually from CDF and D\O\ in Run 1 for the different
decay channels, and the combined results~\cite{exp_xsec}.  The
measurements, within their $\sim 30\%$ uncertainties (dominated by the
statistical component), are consistent with SM predictions.  In Run 2,
a precision of $10\%$ is believed achievable with only 1~fb$^{-1}$ of
data.  Many other factors will then limit the measurement, mostly from
calculation of the total acceptance (lepton and $b$-tagging
efficiencies, event generator systematics, jet energy scale and
luminosity measurement uncertainty, amongst others).  Prospects for
reducing these various components are addressed as needed in
Sec.~\ref{sec:properties}.

\subsection{Single top production}
\label{subsec:production:SM_EW}

Single top quark production cannot occur in flavor-conserving QCD, so
it probes the charged-current weak interaction connecting top to the
down-type quarks, with amplitudes proportional to the CKM matrix
element $V_{tq}$ ($q=d,s,b)$.  This interaction has a vector minus
axial-vector ($V-A$) structure because only the left-chiral component
of fermions participate in the $SU(2)$ gauge interaction.  Also due to
the weak interaction, single top quarks are produced with nearly
$100\%$ polarization, which serves as a test of the $V-A$ structure.

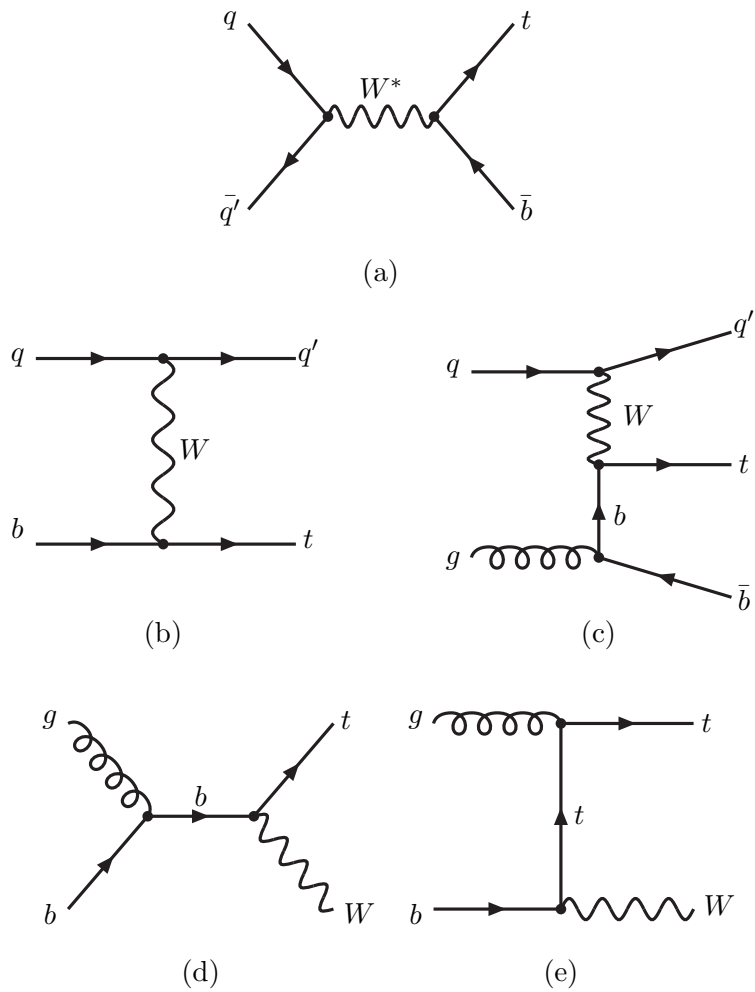
\begin{figure}
\begin{center}
        \begin{picture}(90,80)(0,0)
        \SetOffset(50,40)
        \SetWidth{1.2}
        \ArrowLine(-20,0)(-50,-35)
        \ArrowLine(-50,35)(-20,0)
	\Vertex(-20,0){2}
        \Photon(-20,0)(20,0){4}{4}
	\Vertex(20,0){2}
        \ArrowLine(50,-35)(20,0)
        \ArrowLine(20,0)(50,35)
        \Text(-57,33)[b]{$q$}
        \Text(-57,-30)[t]{$\bar{q^\prime}$}
        \Text(0,7)[b]{$W^\ast$}
        \Text(55,33)[b]{$t$}
        \Text(55,-30)[t]{$\bar{b}$}
        \Text(0,-65)[b]{(a)}
        \end{picture}
\end{center}
\vskip 0.8 cm
\begin{center}
        \begin{picture}(90,110)(0,0)
        \SetOffset(50,55)
        \SetWidth{1.2}
        \ArrowLine(-48,35)(0,35)
        \ArrowLine(-48,-35)(0,-35)
	\Vertex(0,-35){2}
        \Photon(0,35)(0,-35){4}{4}
	\Vertex(0,35){2}
        \ArrowLine(0,35)(50,35)
        \ArrowLine(0,-35)(50,-35)
        \Text(-55,32)[b]{$q$}
        \Text(-55,-33)[b]{$b$}
        \Text(12,-3)[b]{$W$}
        \Text(55,32)[b]{$q^\prime$}
        \Text(55,-30)[t]{$t$}
        \Text(0,-75)[b]{(b)}
        \end{picture}
\hskip 2.5 cm
        \begin{picture}(90,110)(0,0)
        \SetOffset(50,50)
        \SetWidth{1.2}
        \Gluon(-48,-35)(0,-35){4}{4}
        \ArrowLine(-48,35)(0,35)
	\Vertex(0,35){2}
        \Photon(0,0)(0,35){4}{4}
	\Vertex(0,0){2}
        \ArrowLine(0,-35)(0,0)
	\Vertex(0,-35){2}
        \ArrowLine(0,35)(50,50)
        \ArrowLine(0,0)(50,0)
        \ArrowLine(50,-50)(0,-35)
        \Text(-55,33)[b]{$q$}
        \Text(-55,-33)[t]{$g$}
        \Text(55,48)[b]{$q^\prime$}
        \Text(55,-3)[b]{$t$}
        \Text(15,15)[b]{$W$}
        \Text(8,-23)[b]{$b$}
        \Text(55,-55)[b]{$\bar{b}$}
        \Text(0,-70)[b]{(c)}
        \end{picture}
\end{center}
\vskip 1.2 cm
\begin{center}
        \begin{picture}(90,80)(0,0)
        \SetOffset(50,40)
        \SetWidth{1.2}
        \Gluon(-50,35)(-20,0){4}{4}
        \ArrowLine(-50,-35)(-20,0)
	\Vertex(-20,0){2}
        \ArrowLine(-20,0)(20,0)
	\Vertex(20,0){2}
        \Photon(20,0)(50,-35){4}{4}
        \ArrowLine(20,0)(50,35)
        \Text(-57,33)[b]{$g$}
        \Text(-57,-33)[t]{$b$}
        \Text(0,4)[b]{$b$}
        \Text(55,33)[b]{$t$}
        \Text(60,-33)[t]{$W$}
        \Text(0,-65)[b]{(d)}
        \end{picture}
\hskip 1.5 cm
        \begin{picture}(90,80)(0,0)
        \SetOffset(50,40)
        \SetWidth{1.2}
        \Gluon(-48,35)(0,35){4}{4}
        \ArrowLine(-48,-35)(0,-35)
	\Vertex(0,-35){2}
	\Vertex(0,35){2}
        \ArrowLine(0,-35)(0,35)
        \Photon(0,-35)(50,-35){4}{4}
        \ArrowLine(0,35)(50,35)
        \Text(-55,33)[b]{$g$}
        \Text(-55,-33)[t]{$b$}
        \Text(7,-3)[b]{$t$}
        \Text(55,32)[b]{$t$}
        \Text(60,-30)[t]{$W$}
        \Text(0,-65)[b]{(e)}
        \end{picture}
\end{center}
\caption{Leading order Feynman diagrams for electroweak production of
  single top quarks: (a) $s$-channel, (b,c) $t$-channel, and (d,e)
  associated production with a $W$.}
\label{fig:singletop}
\end{figure}

Figure~\ref{fig:singletop} shows the three different ways a hadron
collision can produce top quarks singly.  The process $q\bar{q}\to
t\bar{b}$ via a virtual $s$-channel $W$ boson probes the top quark
with a timelike $W$ boson, $q^2>(m_t+m_b)^2$, while the $W$-gluon
fusion ($t$-channel) processes involve a spacelike $W$ boson, $q^2<0$.
These production mechanisms are thus complementary, as they probe the
charged-current interaction, in different $q^2$ regions.  In the third
process, {\it associated-production}, the $W$ is real and produced in
association with the top quark.

The cross sections for all three processes are proportional to
$|V_{tb}|^2$. Therefore, measuring the single top quark production
cross section provides a direct probe of $|V_{tb}|$ and the weak $tbW$
vertex in general (we discuss $V_{tb}$ in detail in
Sec.~\ref{subsec:properties:lifetime}).  Each process can be affected
by new physics in a different way.  It is therefore important to
observe and study each process separately, to the extent allowed by
the overlap of the signatures.  Studies show that the $s$- and
$t$-channels should be observed at the Tevatron in Run 2 with a small
data sample of only a few fb$^{-1}$.  The associated production
process, however, is smaller in the SM and will be observed only at
the LHC.  As we shall see, the observation of single top is even more
challenging than $t\bar{t}$.  Not only are the cross sections smaller,
but the final state signatures suffer from larger background due to
the less distinctive topology of fewer high-$p_T$ jets, leptons,
b-quarks, and \mET.

It is interesting to note that $p\bar{p}\to t\bar{b}\to W b\bar{b}$ is
a significant background to the SM Higgs search channel $p\bar{p}\to
W^+H$; $H\to b\bar{b}$.  Top quarks, produced either singly or in
pairs, will generally be a background to a host of other channels of
possible new physics.  So even if we are satisfied of the SM
properties of top, we must strive for exacting precision in modeling
top for the sake of searches for new phenomena.

\subsubsection{Single top production in the $s$-channel}

The purely EW $s$-channel single-top process is shown in
Fig.~\ref{fig:singletop}(a).  As this arises from initial-state
quarks, where the PDFs are well-known, the hadronic cross section has
relatively small PDF uncertainty.  The NLO
calculations~\cite{Smith:1996ij,Mrenna:1997wp,Harris:2002md} show
that, for both the Tevatron and the LHC, there is only a relatively
small residual dependence on the scales $\mu_f,\mu_r$, about $\pm
2\%$.  Resummation effects are small, of the order of
$3\%$~\cite{Mrenna:1997wp} and Yukawa corrections (loops involving the
Higgs sector fields) are negligible ($<1\%$) at both colliders.  The
cross section does change, however, by about $\mp 10\%$ at both the
Tevatron and LHC if $m_t$ is varied by $\pm 5$~GeV.  Thus, 1-2~GeV
precision in $m_t$ would be desireable to avoid increasing the
theoretical uncertainty further.  Because the cross section is
potentially so precisely known, this channel may provide the best
direct measurement of $|V_{tb}|$ at the Tevatron (see
Sec.~\ref{subsec:properties:lifetime}).

In Run 1, the cross section was predicted to be about $0.70\pm
0.04$~pb.  This is roughly 8 times smaller than $\sigma_{t\bar{t}}$,
and suffers from comparatively larger backgrounds.  An increase of
only about $30\%$ is expected in Run 2, while an additional factor of
24 is expected for the LHC.  Table~\ref{tab:txsec} shows the results
of the fully differential NLO calculations~\cite{Harris:2002md}.  In
spite of the small cross section, as we discuss below, both CDF and
D\O\ started the search for single top already during Run 1, both to
establish the technique that will bear fruit in Run 2, and on the
chance that new physics increases this cross section greatly beyond SM
expectations.

\subsubsection{Single-top production in the $t$-channel}

The $W$-gluon fusion cross section is illustrated by the Feynman
diagrams in Fig.~\ref{fig:singletop}(b,c). These diagrams are closely
related: diagram (b) shows the hard matrix element to calculate when
the initial parton is treated with a $b$ quark density ($b$ in the
proton sea arises from splitting of virtual gluons into nearly
collinear $b\bar{b}$ pairs); diagram (c) is relevant if the initial
parton is treated as a gluon, and the extra final-state $b$ quark is
typically required to appear at large (experimentally observable)
$p_T$.  Calculation is less precise than for the $s$-channel because
it involves gluon or $b$ quark PDFs, which have relatively large
uncertainties.  In general the inclusive cross section with resummed
logarithms predicts the total single top rate more precisely.  On the
other hand, an exclusive calculation using gluon densities and a
finite transverse momentum `incoming' bottom might in some cases give
better kinematic distributions.  Recent literature~\cite{bpartons} has
highlighted this and corrected some improper uses of $b$ parton
densities, in the context of Higgs boson production.  There, some
factorization scale issues have been shown to be important, which
eventually must be applied to the single top case.

The final state in this channel is $Wbq$, with an occasional
additional $\bar{b}$ antiquark: $\sim 75\%$ of the total cross section
occurs for $p_T(\bar{b})<20$~GeV~\cite{Stelzer:1998ni}, too low to be
observed.  Absence of the additional $b$-jet helps differentiate this
process from the $s$-channel, but the primary distinction is the
additional light quark jet. This will typically be emitted at large
rapidity, very forward in the detector, where most hard QCD events do
not emit jets.  This is sometimes known as a {\it forward tagged jet}.

This channel benefits from a larger production rate compared to the
$s$-channel.  At the Tevatron it is about a factor of three larger,
while at the LHC it is about a factor of 23.  The NLO cross
section~\cite{Bordes95,Stelzer97,Tait:1997fe} retains a somewhat
larger scale dependence than in the $s$-channel case, about $5\%$ at
both the Tevatron and the LHC, but this is still quite good.  If the
top mass is changed by $\pm 5$~GeV, the cross section changes by about
$\mp 8\%(\mp 3\%)$ at the Tevatron (LHC), so its dependence on $m_t$
is comparatively smaller and likely not the limiting factor in
theoretical uncertainty.  The Yukawa corrections are also small,
$\approx 1\%$. The fully differential NLO cross sections for the
Tevatron and LHC are listed in
Table~\ref{tab:txsec}~\cite{Harris:2002md}.

\subsubsection{Associated production channel}

Associated production of single top, $tW$, shown in
Figs.~\ref{fig:singletop}(d,e), proceeds via an initial $gb$ pair,
which makes the cross section negligible at the Tevatron.  However, at
the LHC it contributes about $20\%$ of the total single top cross
section.  Like the $t$-channel case, one of the initial partons is a
$b$ quark.  However, unlike the $t$-channel, the rate of this process
scales like $1/s$.  This, combined with the higher $x$ values needed
to produce a top and a $W$ and correspondingly scarcer quark parton
densities, leads to a cross section about five times smaller than that
of the $t$-channel, despite the fact that associated production is
order $\alpha_s\alpha_W$ rather than $\alpha_W^2$ (the ratio of
strengths is ${\alpha_s\over\alpha_W}\approx 10$).  This cross section
has been calculated only at LO, with a subset of the NLO calculations
included~\cite{Heinson97}; it's relative unimportance make a full NLO
calculation not likely necessary.  Cross section uncertainty is
$\approx 10\%$ from PDFs and $\approx 15\%$ from scale variations.
The cross section at the LHC in the SM is $62$~pb with a total
uncertainty of $\sim 30\%$ (see also Table~\ref{tab:txsec}).

\subsubsection{Experimental status and prospects}

Combining the $s$- and $t$-channel cross sections, the total
single-top production rate is about $40\%$ of $\sigma_{t\bar{t}}$ at
both Tevatron and LHC. Observing singly produced top quarks is more
difficult than those pair-produced, because the final state of
single-top events is not as rich in particle content or pole
structure.  Experimental searches for single top have to take into
account subtle kinematical differences between the relatively larger
backgrounds and the various single-top production channels.  In all
cases, at least one $W$ boson and one $b$ jet are present in the final
state.  To suppress backgrounds from QCD, one is forced to focus on
the leptonic $W$ decay sub-channels, just as the all-hadronic
$t\bar{t}$ channel is difficult at the Tevatron and an extreme
proposition at the LHC; and of course $b$-tagged events.  Therefore,
the starting sample for these searches requires a single high-$p_T$
isolated lepton, large \mET and a $b$-tagged jet.  The challenge is to
understand very precisely the rate and kinematics of all processes
that contribute to the ``$W+b$+jets sample".  Only at that point, and
with enough data that a statistically significant signal can be
extracted, can a credible claim of single-top observation be made.  We
now briefly discuss the searches made at the Tevatron in Run 1 and the
prospects for Run 2 and the LHC.

\noindent\underline{\it Run 1 searches:}

The CDF and D\O\ experiments have searched for each of the potentially
accessible $s$- and $t$-channel signatures separately, and CDF has
also performed a combined search, which looked for single top in the
$W$+jets sample, with the $W$ decaying leptonically into $e$ or $\mu$,
and allowing up to 3 jets.  The invariant mass of the lepton, \vmET
and highest-$p_T$ jet must lie between 140 and 210~GeV, bracketing the
top mass.  This was followed by a likelihood fit to $H_T$, the scalar
$p_T$ sum of all final state objects seen in the detector.  This
distribution is on average softer for non-top QCD backgrounds and
harder for $t\bar{t}$ production, with single top falling somewhere
between.  The limit extracted by this technique is $\sigma(p\bar{p}\to
t+X)<14$~pb at $95\%$ C.L.~\cite{CDFsingletop}.

For the search that separates $s$- from $t$-channel production, CDF
took advantage of $b$-tagging using displaced vertices, and of the
fact that usually only one $b$-tagged jet can be expected in the
$t$-channel case.  This is because the $\bar{b}$ tends to be collinear
with the initial gluon, therefore having too-low $p_T$ to be observed.
The single and double-tagged events in the $W$+2-jet samples were
reconstructed separately and subjected to a likelihood fit.  The
resulting limits~\cite{CDFsingletop} are $\sigma_{s-{\rm chan}}<18$~pb
and $\sigma_{t-{\rm chan}}<13$~pb.

The D\O\ experiment used a neural network trained differently for the
different channels, and considered tagged and untagged events (tagging
for D\O\ was done by associating non-isolated soft muons to
semileptonic $b$-decays).  The limits obtained are~\cite{D0singletop}:
$\sigma_{s-{\rm chan}}<17$~pb and $\sigma_{t-{\rm chan}}<22$~pb.
These limits are about an order of magnitude above the expected SM
values (see Table~\ref{tab:txsec}), but still useful as an
establishment of technique, and to rule out major deviations due to
new physics.

The backgrounds in these searches arose mainly from $W$+jets, QCD
multijets and $t\bar{t}$, with a $S$:$B$ ratio in the range of 1:10 to
1:25, depending on channel and the strictness of event selection.  It
proved cruicial to use $b$-tagging to reduce the background from QCD
multijets (only fakes remained) and from $W$+jets (principally only
$W$+heavy-flavor remained).

\noindent\underline{\it Run 2 and LHC plans:}

At Tevatron Run 2 and the LHC, emphasis will be on the slight
differences in kinematic distributions between the various signal and
background processes to extract the signal in each of the three
channels.  Useful variables include jet multiplicity, event invariant
mass, reconstructed top invariant mass, invariant mass of all jets,
$E_T$ of the jets (including forward jets), $H_T$, and others.
Sophisticated pattern-recognition techniques, such as neural networks
with these or similar inputs, will play a large role.  Such techniques
are now being perfected in order to conduct these searches with better
precison.

Run 2 with only 2~fb$^{-1}$ should be able to achieve $20-30\%$
accuracy for the $s$- and $t$-channel cross section.  At the LHC, the
$t$-channel, the highest yield of the three, is expected to give the
most precise cross section and thus $|V_{tb}|$ measurement.  A $S$:$B$
of about 2:3 should be reached, with statistical uncertainty of
$1-2\%$.  For the $s$-channel at LHC, requiring 2 high-$p_T$
$b$-tagged jets and no other jets in the event yields
$S$:$B\approx$~1:12 and statistical uncertainty of about $6\%$.  For
the associated production channel (accessible only at LHC) to maximize
signal significance, hadronic decays of the $W$ may be included in the
search by constraining a two-jet invariant mass to be close to $M_W$.
This requirement, together with the higher jet multiplicity in the
event, helps reduce backgrounds.  Simulations predict
$S$:$B\approx$~1:4 and statistical uncertainty of about $4\%$.

It is not easy to estimate firmly the systematic uncertainties in
these measurements.  Luminosity alone can contribute at the level of
$5\%$ or more.  Further work on this issue must build on the
experience gained at the Tevatron.

\begin{table}
\def~{\hphantom{0}}
\caption{Single top quark production cross sections (pb).}
\label{tab:txsec}
\vspace{2mm}
\begin{tabular}{|c|c|c|c|c|}
\toprule
Process & Tevatron Run 1 & Tevatron Run 2 & LHC ($t$) & LHC ($\bar{t}$) \\
\colrule
\hline
$\sigma_{s-chann}^{NLO}$ 
& $0.380\pm 0.002$ & $0.447\pm 0.002$
& $6.55\pm 0.03$ & $4.07\pm 0.02$ \\
$\sigma_{t-chann}^{NLO}$ 
& $0.702\pm 0.003$ & $0.959\pm 0.002$
& $152.6\pm 0.6$ & $90.0\pm 0.5$ \\
$\sigma_{assoc.}^{LL}$ 
& - & $0.093\pm 0.024$
& $31^{+8}_{-2}$ & $31^{+8}_{-2}$ \\
\hline
\botrule
\end{tabular}
\end{table}

\subsection{Sensitivity to New Physics}
\label{subsec:production:exotics}

Top quark production at hadron colliders, be it $t\bar{t}$ or single
top, is an ideal place to look for new physics.  If there is any new
physics associated with the generation of mass, it may be more
apparent in the top quark sector than with any of the other lighter,
known, fermions.  Many models predict new particles or interactions
that couple preferentially to the third generation and in particular
to the top quark.  These models extend the strong, hypercharge or weak
interactions in such a way that the new groups spontaneously break
into their SM subgroup at some scale: $SU(3)_h\times SU(3)_l\to
SU(3)_C$, $SU(2)_h\times SU(2)_l\to SU(2)_{W}$, and $U(1)_h\times
U(1)_l\to U(1)_Y$, where $h$ represents the third (heavy) generation
and $l$ the first two (light) generations.  As a result, one would
expect production rate and kinematic distributions of the decay
products to differ from the SM predictions.

Here we highlight only a few scenarios, simply to illustrate the rich
ways top production can be affected by physics beyond the SM.  Along
the way we refer the reader to key papers in the vast literature on
this subject.

\noindent\underline{\it Top pair production:}

In $t\bar{t}$ production, it is especially interesting to study the
invariant mass distribution of the top pair, $d\sigma/dm_{t\bar{t}}$,
since it can reveal resonant production mechanisms. Other interesting
kinematical distributions are the angle of the top quark with respect
to the proton direction (Tevatron only) in the center-of-mass
system~\cite{Lane9501260}, and the top quark and $W$ boson $p_T$
spectra. A partial list of new phenomena that can contribute to the
cross section enhancements and to the distortion of the SM kinematical
distributions can be found in
Refs.~\cite{Eichten94,Hill94,Harris99,Casalbuoni95,Holdom95,Popovic00,Appelquist00,Lane9501260}.


One potential source of new physics in $t\bar{t}$ production is SUSY
correction to QCD~\cite{SUSYxsec}, SUSY being one of the leading
candidates for new physics.  The conclusion is that aside from special
regions in MSSM parameter space, the contribution is at most a few
percent correction to the total $t\bar{t}$ rate or the $m_{t\bar{t}}$
spectrum, making it very difficult to detect SUSY this way.

In another scenario, if the top is a composite quark then there would
be effects modifying the cross section, depending on the properties of
the constituents of the top quark.  If these carry color, scattering
proceeds through gluon exchange~\cite{Cho:1994yu,Atwood:1994vm}. If
the light quarks are also composite then $q\bar{q}\to t\bar{t}$ can
proceed directly through the underlying composite interactions, as
well as by QCD gluon exchange~\cite{Eichten:1983hw}.  In either case,
compositeness would result in an enhancement of the $t\bar{t}$ cross
section over the SM value which could manifest itself as an
enhancement in $d\sigma/dm_{t\bar{t}}$ at large $m_{t\bar{t}}$.


Many theories postulate heavy resonances decaying to $t\bar{t}$, such
as technimesons in technicolor models~\cite{Eichten94,Lane95} (e.g.
$gg\to \eta_T\to t\bar{t}$) or other models of strong
EWSB~\cite{Hill94,Casalbuoni95}.  Variants of technicolor theories,
such as topcolor~\cite{Hill:1991at,BSM_TC2} and topcolor-assited
technicolor (TC2)~\cite{BSM_TC2}, hypothesize new interactions, e.g.
mediated by top-gluons or new weak bosons that are specifically
associated with the top quark, that give rise to heavy states:
$q\bar{q}\to g_t\to t\bar{t}$, $q\bar{q}\to Z^\prime\to t\bar{t}$,
etc.  Since $t\bar{t}$ production at the LHC is dominated by $gg$
fusion, color octet resonances ({\it colorons}) could also be
produced~\cite{Simmons97}.  More recently, extra-dimensional theories
propose scenarios in which new scalar bosons have couplings
preferential to the third generation.  Some scenarios in which only
these bosons live in the extra dimensions predict particles very
similar to the topcolor $Z^{\prime}$~\cite{Appelquist00}.

Top quark pair production can be thought of as the modern day
Drell-Yan, probing the ultra-heavy intermediate states predicted by
various models.  Present and future experiments should patiently scan
the $m_{t\bar{t}}$ spectrum for surprises.  CDF and D\O\ have already
in Run 1 searched for narrow vector resonances in $m_{t\bar{t}}$ in
the single lepton channel.  Within the limited statistics of these
samples (63 events, with $S:B\approx$ 1:1 for CDF), no significant
peaks were observed.  Even though the searches were in principle
model-independent, limits on specific models can be extracted.  CDF
finds that the existence of a leptophobic $Z^\prime$ in a TC2 model
with mass $<480$~GeV ($<780~GeV$) can be excluded at $95\%$~C.L. if
its width is $1.2\%(4\%)$ of its mass~\cite{Affolder:2000eu}.  The
D\O\ search excludes $M_{Z^\prime}<560$~GeV at $95\%$~C.L. for
$\Gamma_{Z^\prime}=0.012 M_{Z^\prime}$~\cite{MZ_tt_D0}.  These
searches will continue in Run 2, extending limits considerably, or
perhaps revealing something more interesting.

\begin{figure}
\epsfxsize=300pt 
\epsfbox{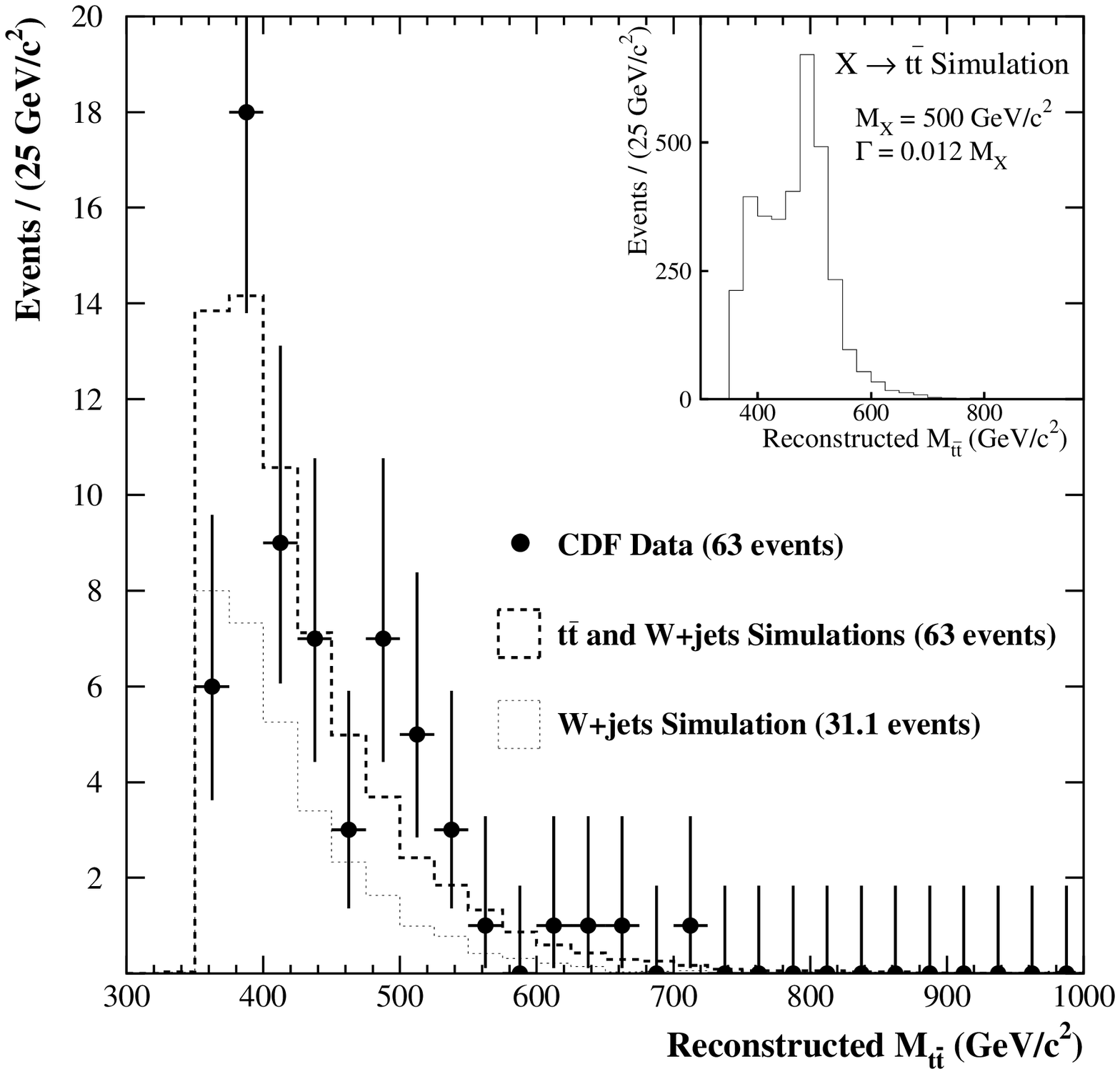}
\caption{The reconstructed $m_{t\bar{t}}$ distribution in the Run 1 
  data from the CDF experiment.}
\label{MTT_CDF}
\end{figure}

Many other kinematical distributions in the top samples were examined
in Run 1~\cite{Abe:1998pm,Top_at_tevatron}, testing consistency with
SM expectations (see e.g.
Refs.~\cite{Frixione:1995fj,Orr97,Lane9501260}).  Within the limited
statistics of the samples, no significant deviations from the SM have
yet been observed.  Nonetheless, some intriguing features, such as
large \vmET and lepton $p_T$, have been noticed in the dilepton
samples~\cite{Barnett:1996zr}.  These could conceivably be attributed
to SUSY production.  However, multi-variable consistency checks do not
show overall significant deviations~\cite{CDFpriv}.  Other samples
that overlap with top, such as the CDF $b$-tagged $W$+jets sample,
show very interesting features, with certain sub-samples containing
soft-lepton tags showing minor deviations from SM expectations. Run 2
data will help decide if these are statistical fluctuations or if some
new physics is hiding in the data.

The LHC could, of course, discover particles with masses larger than
those accessible at the Tevatron.  Studies for the ATLAS experiment
show $5\sigma$ discovery potential curves for $(\sigma\cdot{\cal B})$
v.  $m_{t\bar{t}}$ for a hypothetical narrow
resonance~\cite{Beneke:2000hk}.  Particles as massive as 2~TeV could
be discovered with datasets of 300~fb$^{-1}$ if $\sigma\cdot{\cal
  B}>50$~fb.

\noindent\underline{\it Single top production:}

New physics could also be discerned in single-top production by
introducing new weak
interactions~\cite{Carlson:1993dt,Carlson:1994bg,Hikasa:1998wx,Tait:1997fe,Boos:1999dd};
via loop
effects~\cite{Atwood:1996pd,Simmons97,Li:1997qf,Li:1997ae,Bar-Shalom:1997si};
or by providing new sources of single-top quark
events~\cite{Tait:1997fe,Simmons97,Malkawi:1996fs,Oakes:1997zg,Han:1998tp}.
  
Resonances can also appear in single-top production.  For example, a
new heavy vector boson $W^{\prime\pm}$ or charged scalar $\phi^\pm$,
new $SU(2)$ structure or extra-dimensions, can all contribute
additional diagrams analogous to those in Fig.~\ref{fig:singletop} and
affect the rates and kinematics differently.  The s-channel process
would be particularly sensitive to these states, but the $t$- and
associated production channels are not expected to be affected
significantly~\cite{Tait:1997fe}.  Charged scalars feature in models
with more than one Higgs doublet, such as the MSSM, and in topcolor.
Processes such as $c\bar{b}\to\pi_t^+\to t\bar{b}$ contribute
significantly to the $s$-channel rate (a factor of two enhancement is
possible at the Tevatron and even more at the LHC).  On the other
hand, non-SM flavor-changing neutral currents (e.g. a $Ztc$ vertex)
would be difficult to see in the $s$-channel, while the $t$-channel
would exhibit large effects~\cite{Tait:1997fe}.

Regardless of the specific search for new physics in top quark
production, an important point is that one has to be careful, when
studying kinematical distributions, in making event selections
optimized to detect pure SM production that may dilute the effects of
new physics.  For example, a resonance in $t\bar{t}$ production may
distort the summed $E_T$ and sphericity or aplanarity distributions of
candidate events from their SM expectation~\cite{Lane9501260}.

\section{Top Quark Decays}
\label{sec:decay}

The SM predicts ${\cal B}(t\rightarrow bW)>0.998$.  Other decays
allowed in the SM are not only rare, but also mostly too difficult to
disentangle from backgrounds to be observed in the foreseeable future.
Nevertheless, one must try to be sensitive to all conceivable
signatures of top quark decay, as some can be enhanced by several
orders of magnitude in scenarios beyond the SM, falling within the
LHC's reach.  We first review the SM decays, then discuss
possibilities in the presence of new physics.

\subsection{Standard Model top quark decays}
\label{subsec:decay:SM}

After $t\to bW$~\footnote{Henceforth, we won't distinguish flavor or
  antiflavor whenever the symmetry is obvious. All statements are
  equally valid under charge conjugation.}, the next most likely modes
are the off-diagonal CKM decays $t\to Ws,Wd$.  Together with $t\to
WbZ$, these are the only ones allowed in the SM at tree level and
discussed in Sec.~\ref{subsubsec:decay:SM_CC}\footnote{The radiative
  decays $t\to Wbg$ and $t\to Wb\gamma$ are common, but do not offer
  any fundamental new insight unless the branching fractions turn out
  to be significantly different from the SM predictions (0.3 and
  $3.5\times 10^{-3}$ respectively, for $E_{g/\gamma}>10$ GeV at LHC).
  These channels are generally treated inclusively with $t\to Wb$.}.
Flavor-changing neutral currents (FCNC) decays, $t\to X^0q$, where
$X^0=g,\gamma,Z,H$ and $q=c,u$, are loop induced and highly suppressed
by the GIM mechanism~\cite{GIM}. Branching fractions are typically
${\cal O}(10^{-13})$.  We discuss these in
Sec.~\ref{subsubsec:decay:SM_FCNC}.

\subsubsection{Charged current decays}
\label{subsubsec:decay:SM_CC}

In the SM, $t\to Wb$ is described purely by the universal $V-A$
charged-current interaction.  Being on-shell, however, the $W$ boson's
helicity in top decays is very different from that in the decays of
any other quark, where the $W$ is highly virtual.  The amplitude for
positive helicity $W^+$ boson is suppressed by a chiral factor
$\frac{m^2_b}{M^2_W}$, so the $W$ helicity is a superposition of just
the zero and negative helicity states.  At tree level in the SM, the
fraction ${\cal F}_0$ of the longitudinal (zero helicity) $W$ bosons
in the top rest frame is~\cite{SM_CC_W_helicity_theory,Dalitz:wa}:
\begin{equation}
{\cal F}_0=\frac{m^2_t/M^2_W}{1+m^2_t/M^2_W}=0.701 \pm 0.016
\end{equation}
for $m_t\gg M_W$.  The large top mass exposes the longitudinal mode of
the $W$, so precise measurement of ${\cal F}_0$ serves as a stringent
test of the SM.  To this end, CDF analyzed the lepton $p_T$ spectrum
in $t\bar{t}$ single lepton final states in Tevatron Run 1.  Assuming
a pure $V-A$ coupling, they obtained ${\cal
  F}_0=0.91\pm0.37{\rm(stat.)}\pm 0.13{\rm(syst.)}$, consistent with
the SM~\cite{PDG2002,SM_CC_Vtb_CDF}.  The statistical uncertainty will
be reduced by an order of magnitude in Run 2, and to a negligible
level at the LHC.  Improvement in the systematic uncertainty has yet
to be estimated, but should be better than a factor of 2.

Variables like the angle between the lepton and its parent $W$
direction in the top rest frame depend on the $W$ helicity.  Such
variables as $M_{\ell b}$ can therefore be used to estimate the
relative $W$ helicity fractions, and thus the $V+A$ component in top
decay.  CDF's Run 1 analysis gives $f(V+A) =
-0.21^{+0.42}_{-0.25}{\rm(stat.)}\pm 0.21{\rm(syst.)}$
(preliminary)~\cite{SM_CC_Vtb_CDF_2}.

The ``radiative'' decay $t\to WbZ$ has been suggested~\cite{SM_CC_WbZ}
as a sensitive probe of the top quark mass, since the measured value
of $m_t$ makes this decay close to threshold.  The branching fraction
varies by a factor of $\sim 3$ within the current experimental
uncertainty of $\sim 5$ GeV on $m_t$, but is in the range ${\cal
  O}(10^{-7} - 10^{-6})$, well beyond the sensitivity of the LHC or a
LC.

\subsubsection{Neutral current decays}
\label{subsubsec:decay:SM_FCNC}

With current experimental input, the SM predicts ${\cal B}(t\to cg)
\sim 4\times 10^{-13}$, ${\cal B}(t\to c\gamma)\sim 5\times 10^{-13}$,
and ${\cal B}(t\to cZ)\sim 1\times 10^{-13}$~\cite{SM_FCNC_BRs}.
While ${\cal B}(t\to cH^0)$ depends on $M_{H^0}$, it also cannot
exceed $\sim 10^{-13}$.  These are all well below the detection limits
of even the LHC or a LC~\cite{SM_FCNC_detection}.  Direct searches for
FCNC decays by CDF have set limits of ${\cal B}(t\to c\gamma) + {\cal
  B}(t\to u\gamma) <0.032$ and ${\cal B}(t\to cZ) + {\cal B}(t\to uZ)
<0.33$ at $95\%$ C.L.~\cite{SM_FCNC_CDF}.  These limits are dominated
by statistical uncertainties, and are expected to improve by up to a
factor 10 following Tevatron Run 2.  The LHC experiments have also
estimated their $5\sigma$ discovery reach for these processes.  Given
a 100~fb$^{-1}$ data sample, the minimum branching fractions
accessible to ATLAS and CMS are in the vicinity of $2\times 10^{-4}$
for both $t\to Zq$ and $t\to\gamma q$~\cite{Beneke:2000hk}.

Rates are smaller still for $t\to cX^0_iX^0_j$.  Such FCNC decays can
be significantly enhanced, however, in various scenarios beyond the
SM.

\subsection{Top quark decays beyond the Standard Model}
\label{subsec:decay:BSM}

Many channels emerge to compete with top quark SM decays in the
presence of new physics.  Extended Higgs sectors, alternative
mechanisms for EWSB and mass hierarchies among supersymmetric
particles all attach special significance to the top quark.  We first
consider minimal extensions to the SM Higgs sector without invoking
any new symmetries.  Special implications within the framework of the
MSSM are dealt with following that, together with other scenarios
suggested by SUSY.  Finally, we examine topcolor-assisted technicolor
(TC2).

\subsubsection{Decays with an extended Higgs sector}
\label{subsubsec:decay:BSM_2HDM}

The SM Higgs sector consists of a single complex scalar doublet.  The
single, neutral scalar Higgs boson that arises after EWSB does not
affect top decays in any measureable way.  However, with the addition
of a second Higgs doublet comes charged Higgs states.  If
kinematically allowed, $t\to bH^\pm$ can have a significant branching
fraction.  This is important not merely because a richer Higgs sector
is experimentally allowed, but because it is in fact required by some
of the leading candidates for new physics.  The simplest extension is
to two complex scalar doublets, generically called two-Higgs double
models (2HDM).  In this case, EWSB results in five physical Higgs
bosons: two neutral scalars ($h,H$), a neutral pseudoscalar ($A$), and
a pair of charged scalars ($H^\pm$).  Two new parameters enter at tree
level, usually taken to be $M_{A}$ or $M_{H^\pm}$, and
$\tan{\beta}\equiv\frac{v_2}{v_1}$, where $v_i$ are the vacuum
expectation values of the Higgs fields $\phi_i$ ($i=1,2$).  Both
charged and neutral Higgs boson can appear in tree-level top decays,
the latter implying FCNCs.

\noindent\underline{\it Decays to charged Higgs bosons:}

Among a few variants of the two-Higgs-doublet models (2HDM) is the
``Type 2'' model, where one doublet couples to up-type fermions and
the other to down-type.  This is required, for example, in the
MSSM~\cite{BSM_2HDM_type2}.

If $M_{H^\pm}<m_t-m_b$, then 
\begin{equation}
\Gamma(t\to H^+b) \propto 
(m^2_t\cot^2{\beta}+m^2_b\tan^2{\beta})(m^2_t+m^2_b-M^2_{H^\pm})+4m^2_t m^2_b
\end{equation}
at tree level.  For fixed $M_{H^\pm}$, this function is symmetric in
$\log(\tan{\beta})$ about a minimum at
$\tan{\beta}=\sqrt{\frac{m_t}{m_b}}$.  For given $\tan{\beta}$, the
partial width decreases as $M_{H^\pm}$ increases.  If one ignores
fermion masses except when they are multiplied or divided by
$\tan{\beta}$, then in the diagonal CKM approximation the fermionic
decay partial widths are given by
\begin{equation}
\Gamma(H^+\to U\bar D) = \frac{N_c g^2 M_{H^+}}{32\pi M^2_W} 
(m^2_U\cot^2{\beta} + m^2_D\tan^2{\beta}) \; ,
\end{equation}
where $U[D]$ is an up-[down-] type fermion and $N_c=1[3]$ for leptons
[quarks].  With the current experimental lower limit of
$M_{h}>91.0$~GeV and $M_{A}>91.9$~GeV at
$95\%$~C.L.~\cite{BSM_MSSM_h0_LEP}, bosonic decays $H^\pm\to W^\pm
h,W^\pm A$, are kinematically suppressed for $M_{H^\pm}<m_t-m_b$.

Thus, for $\tan{\beta}>1$, $H^\pm\to\tau\nu_\tau$ is the dominant
decay channel. If $\tan{\beta}<1$, the decay depends on $M_{H^\pm}$:
for $M_H^\pm\approx 100$~GeV, $H^\pm\to cs$ and $H^\pm\to bc$ compete
more or less evenly (CKM suppression due to $|V_{cb}| \ll |V_{cs}|$ is
offset by the stronger $H^\pm$ coupling to $b$ relative to $s$); but
as $M_H^\pm$ is increased beyond 120 GeV, weight gradually shifts to
$H^\pm\to Wbb$ via a virtual top quark.  Strategies for $H^\pm$
searches therefore depend on $M_{H^\pm}$ and $\tan{\beta}$.  Searches
for $e^+e^-\to H^+H^-$ at LEP constrain $M_{H^\pm}>78.6$~GeV at
$95\%$~C.L.~\cite{BSM_2HDM_H+_LEP}, while the CLEO experiment has set
a limit of $M_{H^\pm}>(244+63/(\tan{\beta})^{\frac{1}{3}})$~GeV at
$95\%$~C.L. from the inclusive measurement of $b\to s\gamma$
\cite{BSM_2HDM_H+_CLEO}.

\begin{figure}
\epsfxsize=300pt 
\epsfbox{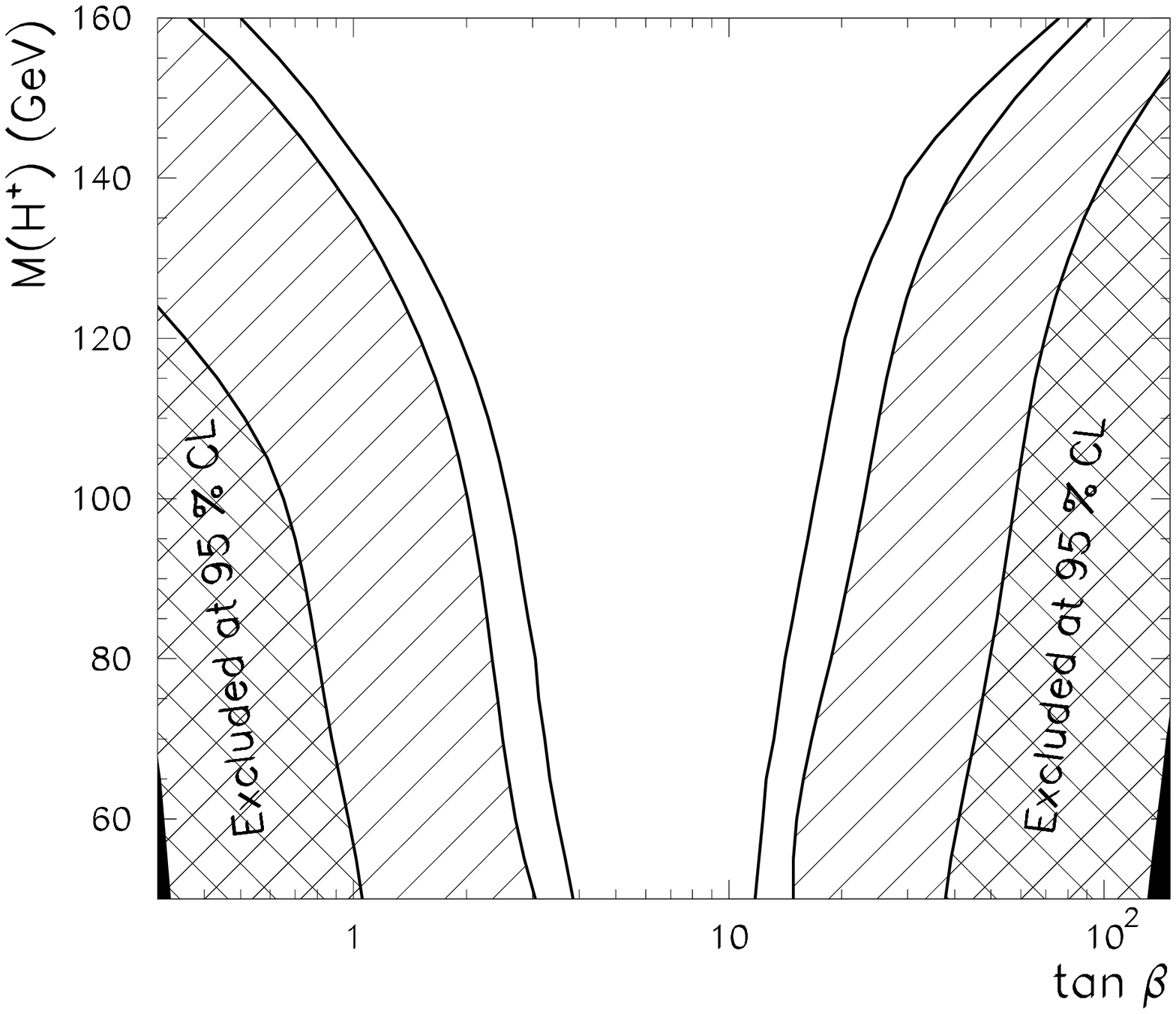}
\caption{The $95\%$~C.L. exclusion boundaries in the [$M_{H^+},\tan{\beta}$] 
  plane from the D\O\ Run 1 ``disappearance search'' for $t\to bH^\pm$
  (double hatched).  Also shown are Run 2 projections if the
  probability of experimental observations continues to peak at the SM
  prediction: 2 fb$^{-1}$ (single hatched), and 10 fb$^{-1}$
  (unhatched).  The modeling is based on leading-order calculations.
  More recent results from LEP have excluded $M_{H^+}<78.6$ GeV at
  $95\%$~C.L.}
\label{fig:BSM_2HDM_H+_tevatron}
\end{figure}

By itself, an extended Higgs sector does not significantly alter
$\sigma_{t\bar{t}}$ at hadron colliders.  One looks instead for either
the appearance of $t\to H^\pm b$ signatures or, indirectly,
disappearance of the SM $t\to Wb$ signatures. For the latter, one
assumes ${\cal B}(t\to H^\pm b) + {\cal B}(t\to Wb)=1$.  Both CDF and
D\O\ conducted searches for $t\to H^\pm b$ in $p\bar{p}\to t\bar{t}$
events in Run 1~\cite{BSM_2HDM_H+_D0,BSM_2HDM_H+_CDF}.
Figure~\ref{fig:BSM_2HDM_H+_tevatron} shows the D\O\ results from
their disappearance search together with projections for Run 2.

The direct searches focused on $H^\pm\to\tau\nu$.  With good $\tau$
identification capability, this can yield the strongest results,
albeit limited to $\tan{\beta}>1$, where the process has a large
branching fraction.  Combinations of different methods and of data
from the two experiments may indeed eventually give stronger
constraints.  As expected, searches are more difficult in the region
around $\tan{\beta}=\sqrt{\frac{m_t}{m_b}}$, where $t\to bH^\pm$ is
highly suppressed.  Searches for $H^\pm\to cs,cb$ are made more
challenging by overlap with the SM decay $t\to Wb\to q_1q_2b$.
However, a dijet invariant mass peak between 110 GeV and 130 GeV
corresponding to $M_{H^\pm}$ is a viable signal for Tevatron Run 2 and
LHC.  For $M_{H^\pm}>130$ GeV, $t\to bH^\pm\to Wbbb$ may offer cleaner
signatures, but ${\cal B}(t\to bH^\pm)$ decreases rapidly with
increasing $M_{H^\pm}$.  Increased statistics from Run 2 and the LHC
will push the exclusion contour wings asymptotically closer (see
Fig.~\ref{fig:BSM_2HDM_H+_tevatron}) - or perhaps the process will be
observed.  The exclusion boundaries in the [$M_{H\pm},\tan\beta$]
plane roughly follow contours of constant ${\cal B}(t\to bH\pm)$.
Thus, $95\%$~C.L. upper limits on ${\cal B}(t\to bH\pm)$ for
$\tan\beta>1$ (where $H\pm\to\tau\nu$ dominates) are 0.36 from D\O\ 
and 0.5-0.6 from CDF.  The disappearance search result from D\O\ can
be interpreted as ${\cal B}(t\to bH\pm)<0.45$ at $95\%$~C.L.,
irrespective of $\tan\beta$ except in the region where $H\pm\to Wbb$
is the dominant decay mode (i.e. when $\tan\beta<1$ and
$M_{H\pm}>125$~GeV).  The corresponding estimate for Run 2 is ${\cal
  B}(t\to bH\pm)<0.11$ at $95\%$~C.L.~\cite{BSM_2HDM_H+_Run2}.

All $H^\pm$ searches hinge on the fact that, unlike for $W^\pm$,
$H^\pm$ fermion couplings are not flavor-blind.  This implies we
should compare the values for $\sigma_{t\bar{t}}$ derived from
different final states, based on the SM assumption of ${\cal B}(t\to
Wb)\approx 1$.  For example, if the dilepton, single-lepton, and
all-jets $t\bar{t}$ final states exhibited differences, it could
indicate significant alternative decay modes to $t\to Wb$.  While less
restrictive in assumptions, this method also yields the least
stringent conclusions.  Tevatron Run 1 data is statistically
insufficient for a meaningful application of this method, but that
will change for Run 2 and the LHC.

\noindent \underline{\it FCNC decays in a 2HDM:}

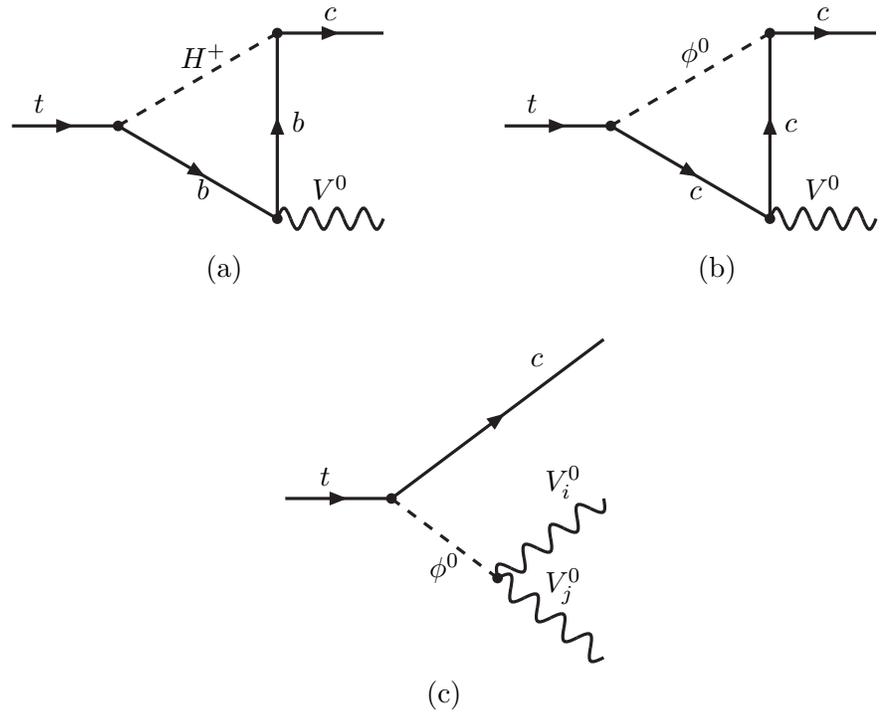
\begin{figure}
  \begin{center}
        \begin{picture}(140,85)(0,0)
        \SetOffset(80,40)
        \SetWidth{1.2}
        \ArrowLine(-80,0)(-40,0)
        \DashLine(-40,0)(20,35){4}
        \ArrowLine(-40,0)(20,-35)
        \ArrowLine(20,-35)(20,35)
	\Vertex(-40,0){2}
	\Vertex(20,35){2}
	\Vertex(20,-35){2}
        \ArrowLine(20,35)(60,35)
        \Photon(20,-35)(60,-35){4}{4}
        \Text(-70,5)[b]{$t$}
        \Text(-8,22)[b]{$H^+$}
        \Text(-8,-28)[b]{$b$}
        \Text(28,-2)[b]{$b$}
        \Text(40,-28)[b]{$V^0$}
        \Text(40,40)[b]{$c$}
        \Text(0,-60)[b]{(a)}
        \end{picture}
\hskip 1.5 cm
        \begin{picture}(140,80)(0,0)
        \SetOffset(80,40)
        \SetWidth{1.2}
        \ArrowLine(-80,0)(-40,0)
        \DashLine(-40,0)(20,35){4}
        \ArrowLine(-40,0)(20,-35)
        \ArrowLine(20,-35)(20,35)
	\Vertex(-40,0){2}
	\Vertex(20,35){2}
	\Vertex(20,-35){2}
        \ArrowLine(20,35)(60,35)
        \Photon(20,-35)(60,-35){4}{4}
        \Text(-70,5)[b]{$t$}
        \Text(-8,22)[b]{$\phi^0$}
        \Text(-8,-28)[b]{$c$}
        \Text(28,-2)[b]{$c$}
        \Text(40,-28)[b]{$V^0$}
        \Text(40,40)[b]{$c$}
        \Text(0,-60)[b]{(b)}
        \end{picture}

        \begin{picture}(120,170)(0,0)
        \SetOffset(40,70)
        \SetWidth{1.2}
        \ArrowLine(-40,0)(0,0)
        \ArrowLine(0,0)(80,60)
        \DashLine(0,0)(40,-30){4}
        \Photon(40,-30)(80,0){4}{4}
        \Photon(40,-30)(80,-60){4}{4}
	\Vertex(-0,0){2}
	\Vertex(40,-30){2}
        \Text(-25,5)[b]{$t$}
        \Text(20,-32)[b]{$\phi^0$}
        \Text(55,50)[b]{$c$}
        \Text(65,0)[b]{$V^0_i$}
        \Text(65,-40)[b]{$V^0_j$}
        \Text(20,-80)[b]{(c)}
        \end{picture}

\end{center}
\caption{One-loop diagrams of $t\to cV^0$ (top) 
  and tree diagrams of $t\to cV^0_iV^0_j$ (bottom) in 2HDM.
  $V^0=\gamma,Z,g$; $\phi^0=h^0,H^0,A^0$.}
\label{fig:BSM_2HDM_FCNC_tcV} 
\end{figure}

FCNC top quark decay rates can be enhanced if one abandons the
discrete symmetry invoked in the Type 2 2HDM to suppress tree-level
scalar FCNCs.  In the more general Type 3 2HDM, fermions are allowed
to couple simultaneously to more than one scalar
doublet~\cite{BSM_2HDM_type3}~\footnote{Low energy limits on FCNCs may
  be explained by tuning of the Yukawa matrices.}.  Single
vector-boson FCNC decays, $t\to cV^0_i$ ($V^0_{i}=\gamma,Z,g$) are
still loop-induced, as shown in Fig.~\ref{fig:BSM_2HDM_FCNC_tcV}(a,b),
but can have branching fractions as large as ${\cal O}(10^{-5})$ even
without any new interactions~\footnote{These branching fractions can
  be enhanced by more than a factor 10 under favorable conditions in
  the MSSM.}.  Double vector-boson FCNC decays, $t\to cV^0_iV^0_j$
also appear at the tree-level (Fig.~\ref{fig:BSM_2HDM_FCNC_tcV}(c)),
and can reach branching fractions of ${\cal
  O}(10^{-5})$~\cite{BSM_2HDM_FCNC}.

With production rates of ${\cal O}(10^3-10^4)$ per year (see
Table~\ref{table:collider_params}), such events could be studied at
the LHC only if they are given high priority in triggering during
high-luminosity running, because suppression of large SM backgrounds
will translate into small signal efficiencies.  At a LC, production
rates are at most ${\cal O}(1-10)$ per year, but low background and
very high ($\sim 90\%$) signal efficiency may make these processes
observable, should they occur.

\subsubsection{Supersymmetric decays of the top quark}
\label{subsubsec:decay:BSM_SUSY}

In SUSY, the large Yukawa coupling of the top quark can lead to large
mass splitting among the superpartners of the third generation
fermions.  The superpartners of the right-handed and left-handed top
quark combine to form mass eigenstates $\tilde{t}_1$ and
$\tilde{t}_2$.  The lightest top squark, $\tilde{t}_1$, can be lighter
than all other squarks, and in fact have mass near $m_t$.  Naturally,
this has implications for possible top decays.  We first address top
SUSY decays under the assumption that ${\cal R}$-parity~\footnote{A
  discrete, multiplicative symmetry defined as ${\cal R}_p \equiv
  (-1)^{3B+L+2S}$, where $B$ is baryon number, $L$ lepton number, and
  $S$ spin.} is conserved.  Afterward, we drop this assumption.

${\cal R}$-parity conservation requires superparticles to be produced
in pairs and forbids decays of the lightest SUSY particle (LSP).  The
LSP is widely assumed to be the lightest neutralino, $\tilde\chi^0_1$
(neutralinos are the sfermion partners of the SM bosons).  Under this
assumption, the most likely top SUSY decay is $t\to\tilde{t}_1
\tilde\chi_1^0$.  Generally, the top squark will decay via
$\tilde{t}_1 \to c \tilde\chi_1^0$ or $b \tilde\chi_1^+$, depending on
the various daughter masses.  In the latter case, $\tilde\chi_1^+ \to
\tilde\chi^0_1 \ell\nu_\ell$ or $\tilde\chi^0_1 q_1\bar q_2$.  The
neutralinos interact only weakly, so generally escape without
detection like neutrinos.

Branching fractions as large as 0.4-0.5 are possible for $t\to
\tilde{t}_1 \tilde\chi_1^0$~\cite{BSM_SUSY_stop_theory}.  In such a
scenario, about one half of $t\bar{t}$ events would have one SM and
one SUSY top decay.  The CDF experiment has searched Run 1 data for
events of this type where the SM top decay proceeds as $t\to Wb\to
\ell\nu_\ell b$ ($\ell=e,\mu$), while the SUSY decay of the other top
proceeds as $t\to \tilde{t}_1 \tilde\chi_1^0 \to b
\tilde\chi_1^+\tilde\chi^0_1 \to bq_1\bar q_2 \tilde\chi^0_1
\tilde\chi^0_1$.  The signal consists of a lepton, \mET, and 4 jets
(including the two $b$-jets): identical to SM single lepton decay, but
differing in $p_T$ and angular distributions.  These depend on the
masses of the particles involved.  Based on the assumptions ${\cal
  B}(\tilde\chi^\pm_1 \to \ell\nu\tilde\chi^0_1)=\frac{1}{9}$, ${\cal
  B}(\tilde{t}_1 \to b \tilde\chi_1^\pm)=1$, and ${\cal B}(t \to
\tilde{t}_1 \tilde\chi_1^0)+{\cal B}(t\to Wb)=1$, the search excluded
${\cal B}(t \to \tilde{t}_1 \tilde\chi_1^0)>0.45$ at $95\%$~C.L. over
most of the kinematically allowed portion of the [$m_{\tilde t_1},
m_{\tilde\chi_1^\pm}$] parameter space for $m_{\tilde\chi^0_1}$ up to
40 GeV~\cite{BSM_SUSY_stop_CDF}.  For larger LSP masses, the
kinematically allowed region shrinks.

The alternative scenario, $t\to \tilde{t}_1 \tilde\chi_1^0 \to c
\tilde\chi_1^0 \tilde\chi_1^0$, is similar in character to the FCNC
decay $t\to cZ \to c\nu\nu$.  The most promising channel is where one
top undergoes the non-SM decay while the other follows the SM.  If the $W$
decays leptonically, then the signal consists of a hight-$p_T$
isolated lepton, substantial $\mET$, and 2 jets, one of which is a
$b$.  The large background from $W(\to\ell\nu)+\ge 2$~jets limits the
search to regions of parameter space where $m_{\ell\vmpT}>M_W$.  If,
on the other hand, the $W$ decays hadronically, then we have 4
high-$p_T$ jets and large \mET for signal.  Backgrounds arise chiefly
from $W(\to \tau \nu)+\ge 3$ jets events where the $\tau$ is
misidentified as a jet, and from $Z(\to\nu\nu)+\ge 4$~jets.
Effectiveness of $b$-tagging is reduced since there is only one
$b$-jet per event.  Sensitivity is further compromised in much of the
[$m_{\tilde{t}_1}, m_{\tilde\chi_1^0}$] parameter space where the the
jet and \mET spectra are soft and/or broad.  Tevatron Run 1 data was
statistically insufficient for this analysis, but that will change in
Run 2.

${\cal R}$-parity violating ($\MissingB{\cal R}{\it p}$) interactions
in the MSSM greatly enhance FCNCs~\cite{BSM_SUSY_RPV_theory}.  Within
a single coupling scheme, either the up-type quarks or the down-type
quarks can avoid these processes, but not both simultaneously.  The
consequences of $\MissingB{\cal R}{\it p}$ have been studied via
measurements of $K^0$-$\bar K^0$, $D^0$-$\bar D^0$ and $B^0$-$\bar
B^0$ mixing, and of ${\cal B}(K^+\to \pi^+\nu\bar\nu$), resulting in
constraints on the $j=1,2$ elements of the $3\times 3\times 3$
$\MissingB{\cal R}{\it p}$ coupling matrix $\lambda^\prime_{ijk}$
($i,j,k$ are generation indices), but leaving the third generation
somewhat unconstrained.  If sleptons lighter than the top quark exist,
then $t_L\to d_{Rk}{\tilde{\ell}}^+_i$ followed by
${\tilde{\ell}}^+_i\to \tilde{\chi}_0 \ell_i$ and $\tilde{\chi}_0 \to
\bar{\nu}_i\bar{b}d_k$ can lead to a fairly clean signature
($\MissingB{\cal R}{\it p}$ implies that the $\tilde{\chi}_0$, assumed
here to be the LSP, is not stable).  Future searches for such signals
will constrain $\lambda^\prime_{i3k}$ ($k\ne 3$).

\subsubsection{Top decays in topcolor-assisted technicolor}
\label{subsubsec:decay:BSM_TC2}

In technicolor theories~\cite{BSM_TC}, EWSB is accomplished by chiral
symmetry breaking of technifermions which transform nontrivially under
a new strong gauge interaction called technicolor (TC).  This yields
correct weak boson masses if the scale of technicolor interactions is
about a TeV.  Fermion masses arise without fundamental scalars, by
invoking an additional, spontaneously broken gauge interaction called
extended technicolor (ETC)~\cite{BSM_ETC_1}.  However, ETC
interactions cannot account for the large mass of the top
quark~\cite{BSM_ETC_2}.

Topcolor-assisted technicolor (TC2) is an attempt to address this
deficiency~\cite{BSM_TC2}.  In the simplest version, the third
generation is assumed to transform with the usual quantum numbers
under strong $SU(3)_h\times U(1)_h$, while the lighter generations
transform identically under a different (weaker) group $SU(3)_l\times
U(1)_l$.  At scales of about 1 TeV, $SU(3)_h\times SU(3)_l$ and
$U(1)_h\times U(1)_l$ spontaneously break down to ordinary color
$SU(3)_C$ and weak hypercharge $U(1)_Y$, respectively.  EWSB is still
driven primarily by TC interactions, but topcolor interactions, felt
only by the third generation quarks (also at a scale near 1~TeV),
generate the very large top quark mass.  ETC interactions are still
required to generate the light fermion masses and a small but
important contribution to the mass of the top quark $m_t^{\rm ETC}$.
The reason for a nonzero $m_t^{\rm ETC}$ is to give mass to the {\it
  top-pions}, the Goldstone bosons of $t,b$ chiral symmetry breaking.

In TC2 models, the $tb\pi^+_T$ coupling is small, but the $tb\pi^+_t$
coupling is large, and the ETC interactions responsible for the small
component of $m_t$ induce mixing between top-pions and technipions.
The consequence is a possibly significant partial width (if
kinematically allowed):
\begin{equation}
\Gamma(t \to \pi^+_tb)=\frac{|\epsilon|^2}{16\pi}
\left(\frac{m_t^{\rm dyn}}{m_t}\right)^2
\frac{(m^2_t-m^2_{\pi_t})^2}{F^2_tm_t},
\end{equation}
where $\epsilon$ is the top-pion component of the technipion mass
eigenstate, $m_t^{\rm dyn}$ the dynamical top quark mass, $m_{\pi_t}$
the technipion mass, and $F_t$ ($\approx 70$ GeV) the top-pion decay
constant.  Short of direct discovery, a precise experimental
determination of $\Gamma_t$ is required to limit the allowed parameter
space in these models.

\section{Top Quark Properties}
\label{sec:properties}

Confirmation of the SM nature of the top quark requires that we
measure all its quantum properties and compare with SM expectations.
Deviations would indicate new physics. 
In this section we describe the status and future expectations of these 
measurements, and the crucial issues in making them.

\subsection{Mass}
\label{subsec:properties:mass}

While the top quark is the least well-studied quark in terms of
quantum properties, its mass, $m_t$, is more accurately known (as a
fraction of its mass) than any other quark. This is also extremely
important, as the top quark's role in SM precision fits is
proportionally more important than any other. This is an artifact of
EWSB and the large value of the top Yukawa coupling, $Y_t$. That $Y_t$
appears to be exactly one has not gone unnoticed.  Proponents of
strong dynamical EWSB argue it that supports this class of theories,
because in general they predict large values of $Y_t$, on the order of
1 or more. On the other hand, it is also generally regarded as support
for SUSY extensions to the SM, which would not be viable unless the
top quark mass were large: the running of $\sin^2{\theta_W}$ could not
be made to fit the data and still allow for gauge coupling unification
otherwise, and EWSB would not occur, since the large value of the top
Yukawa coupling is what drives the coefficient of the Higgs mass term
negative. But large $m_t$ does not point at either class of theories
as the clear favorite.  One is left with the simple suspicion that the
top quark is perhaps connected to new physics on the grounds that
physical parameters of exactly 1, (or 0, etc.) indicate a more
fundamental property underlying $Y_t$.

The impact of $m_t$ elsewhere varies. In $B$ and $K$ physics, many
observables have terms roughly quadratic in $m_t \over M_W$. It was,
in fact, data from $B_0-\bar{B}_0$ mixing in 1987 that first indicated
a heavy top quark. For precision SM EW fits, $m_t$ enters
quadratically in many places as well. Examples are $R_b$, $A_{LR}$,
$\sin^2{\theta_W}$ and the parameter $\rho\equiv{M^2_W\over
  M^2_Z\sin^2(\theta_W)}$. The corrections usually appear as a
multiplicative factor, $1+{3G_Fm^2_t\over 8\sqrt{2}\pi^2}$. The $W$
mass, which is not known nearly as precisely as most of the other
quantities in the EW sector, receives quantum corrections proportional
to $m^2_t$ and $\ln (M_H)$, where $M_H$ is the Higgs boson mass.  This
is usually plotted as $m_t$ v.  $M_W$, overlaid with bands that show
the predicted $M_H$, as in Fig~\ref{fig:mw_mt_mh}.  A ``light'' Higgs
is favored, somewhere around 100~GeV, but with an uncertainty also of
${\cal O}(100)$~GeV.  Unfortunately, as the $M_H$ dependence is only
logarithmic, and in the presence of new physics this fit is not
meaningful unless the new physics is also known precisely, one cannot
draw firm conclusions from these fits. As the precision of $m_t$ and
$M_W$ increases, however, and if a Higgs remains unobserved, the fit
increasingly suggests breakdown of the SM.

The current precision of $B$ and $K$ physics is not good enough to
require better precision in $m_t$ than is available from Tevatron Run
1, but the next generation of $K$ experiments will need $\delta m_t
\simeq 3\% \simeq 5$~GeV, which should be satisfied by Run 2. The EW
precision fits are more demanding. Once the $W$ mass precision reaches
20~MeV at the LHC, $m_t$ must be known within 3~GeV to not limit the
EW precision fit for $M_H$. For a future linear collider, the 6~MeV
precision on $M_W$ must be matched by 1~GeV precision in $m_t$.

Both the LHC and a LC can outperform these goals: at the LHC, $\delta
m_t \simeq 2$~GeV is expected within 1 year of low-luminosity running,
while 1~GeV could be achieved with the $\ell J/\psi$ final state
(discussed shortly) and a larger data set~\cite{Beneke:2000hk}.
Precision of ${\cal O}(100{\rm ~MeV})$ can be obtained at a future
linear collider with a $t\bar{t}$ threshold
scan~\cite{Aguilar-Saavedra:2001rg}, which does not measure the pole
mass and so is not limited by uncertainties of ${\cal
  O}(\Lambda_{QCD})$.

One specific case where super-precision of $m_t$ would be necessary is
if low-energy SUSY is found. In the MSSM, the mass of the lighter
CP-even neutral Higgs boson $h$ is given at the NLO by
\begin{equation}
M^2_h \; = \; 
M^2_Z \, + \, {3G_F\over\pi^2\sqrt{2}} \; m^4_t \; {\rm ln}
\biggl( {M^2_S\over m^2_t} \biggr) \; ,
\end{equation}
where $M^2_S$ is the average of the two top squark squared masses.
Since a LC could measure $M_h$ to about 50~MeV
precision~\cite{Aguilar-Saavedra:2001rg}, $m_t$ would need to be known
to 100~MeV or better to perform meaningful SUSY-EW precision fits.
Ironically, this would require $M_h$ to be known to probably the
four-loop level; only two-loop calculations are currently available.
One is forced to wonder if the requisite improvement in theoretical
precision in that case would be realistic.

We now highlight the principles behind top mass measurements made so
far at the Tevatron. Details and subtleties can be found in e.g.
Refs.~\cite{Tollefson:wt,Beneke:2000hk,Affolder:2000vy,mt_D0_1lep,mt_D0_1lep_new,mt_D0_2lep,mt_CDF}.
The main idea is to compare the observed kinematic features of
$t\bar{t}$ pairs to those predicted for different top quark masses.
While many kinematic variables are sensitive to $m_t$, explicit
reconstruction from the $t\bar{t}$ decay products is an obvious
choice, as long as we understand that it is uncertain to at least
${\cal O}(\Lambda_{QCD})$.  However, more elaborate methods that
attempt to connect many observables simultaneously with the matrix
elements of the production and decay processes on an event-by-event
basis are gradually emerging as a superior alternative.

There are three channels to consider, depending on how the two top
quarks decay: dilepton, single-lepton, and all-hadronic.  Here,
``lepton'' refers to $e,\mu$ only, since the presence of additional
neutrinos in $\tau$ decays severly limits the usefulness of $t\bar{t}
\to \tau X$ channels in the $m_t$ determination.  Thus, the branching
fractions of the three channels are approximately 0.05, 0.30 and 0.44,
respectively.  Signal and background characteristics vary from channel
to channel, so the exact technique used must be tailored accordingly
for each channel.

For direct reconstruction of invariant masses of the two top quarks in
a $t\bar{t}$ candidate event, one needs to know the 4-momenta of the 6
daughters, a total of 24 quantities.  Imagine an ideal $t\bar{t}X$
event with no final state radiation and where the momentum of $X$,
which represents everything recoiling against the $t\bar{t}$ system,
is fully measured.  If the 3-momenta of $n$ of the six final state
objects are directly measured, we have $3n$ measured quantities from
the two top decays.  The masses of the 6 decay products are known
(these can be safely assumed zero), as are the two intermediate $W$
masses.  Although $m_t$ is yet unknown, it must be the same for both
tops in the event. So, we have 9 constraints from
particle masses\footnote{One has to appropriately allow for
  $\Gamma_W$ and $\Gamma_t$.}.  That the $t\bar{t}X$ system carries 
no significant momentum transverse to the beamline gives two additional
constraints\footnote{In general, $x_1\ne x_2 \Rightarrow
  p_z(t\bar{t}X)\ne 0.$}: $\vec{p}_T(t\bar{t}X)=0$.  Thus, a kinematic
mass fit is subject to $(3n+9+2-24)=(3n-13)$ constraints.  For each
leptonic $W$ decay, there is a corresponding neutrino that cannot be
directly observed.
Therefore, $n=6$ for all-hadronic, $n=5$ for single-lepton, and $n=4$
for dilepton events.  Dilepton events are underconstrained ($-1$C),
preventing explicit $m_t$ reconstruction from its daughters, forcing
one to seek alternative means.

In every channel, many factors complicate $m_t$ measurement.  The
observed objects' momenta need to be corrected to remove detector
effects.  The lion's share of the uncertainty in these corrections is
due to jet energy measurements.  Any sampling calorimeter has a
relatively large inherent uncertainty in its absolute energy scale.
Moreover, the detector geometry has non-uniformities such as module
boundaries and gaps or ``cracks'' to allow passage of cables and other
hardware.  Therefore, the response must be carefully mapped as a
function of the physical location of where the jet traversed the
detector.  It is often a non-linear function of jet energy.
Additionally, each element of a calorimeter, or {\it cell}, has a
minimum threshold to register energy.  Reconstruction of jets proceeds
through identification of clusters of (nearly) contiguous cells
registering energy.  These effects usually result in leakage that
needs to be corrected for.  Two other effects come from the nature of
hadron collider events.  In each $t\bar{t}$ hard scattering there is
an associated {\it underlying event} from the proton/antiproton
remnants, that deposits soft energy through the calorimeters. Also, in
high luminosity running, each $t\bar{t}$ event is accompanied by {\it
  multiple interactions}, dominated by soft inelastic $p\bar{p}$ or
$pp$ scattering, that contribute to energy measurement contamination.

Other complications arise, more related to the physics of the
$t\bar{t}$ event itself.  One is that we often find jets that do not
even originate from top decays directly, rather from initial or final
state radiation~\cite{ttj}.  Due to detector segmentation or
limitations in the reconstruction algorithms, two or more jets can get
merged and reconstructed as one.  Sometimes the opposite occurs: a
single jet splits in two due to fragmentation.  Occassionally, a jet
is lost entirely because it travels through an uninstrumented or
poorly instrumented region, such as the beampipe.  These extra or
missing jets result in admission of extraneous solutions into
reconstructed $m_t$ distributions.

Since the all-hadronic channel has a large branching fraction and is
maximally constrained, one might surmise that it would be the best for
measuring $m_t$.  In practice, however, a very large and hard-to-model
QCD multijet background, compounded by the jet measurement issues
mentioned above, leads to relatively large uncertainties.  The top
mass extracted by CDF~\cite{Abe:1997rh} in the all-hadronic channel is
$186.0\pm 10({\rm stat.})\pm 5.7({\rm syst.})$~GeV.  Each event is
required to have six or more jets, and to satisfy several topological
requirements that help improve the signal to background ratio.  Events
were reconstructed to the $t\bar{t}\to W^+bW^-\bar{b} \to q_1\bar{q}_2
b q_3\bar{q}_4\bar{b}$ hypothesis using the six highest $E_T$ jets,
one of which must be $b$-tagged.  This still leaves 30 different
reconstruction combinations.  A kinematic fit constrains each
combination to yield $M_W$ for two jet pairs, equal $t$ and $\bar{t}$
masses, returning a $\chi^2$ value.  The combination with the smallest
$\chi^2$ is chosen.  The resulting ``reconstructed mass" distribution
from the candidate events is then compared, through a likelihood fit,
to templates formed from the right mix of $t\bar{t}$ (from simulation)
and QCD background, the shape of which is extracted from data. The
input $m_t$ is changed and the value that maximizes the likelihood $L$
is the central value of the top mass measurement.  The statistical
uncertainty is determined from the range over which the $-\ln{L}$
increases by $\frac{1}{2}$ unit with respect to its minimum.  An
analysis of the all-hadronic final state recently completed by D\O\ is
similar in spirit, but employs an artificial neural network algorithm
to compensate for a lower $b$-tagging efficiency.  The preliminary
result is $176^{+17.1}_{-13.6}$~GeV.

The ultimate precision achievable in this channel is not expected to
rival that of the single-lepton or dilepton channels but can still be
used in a combined result to help reduce the overall uncertainty.  A
top mass measurement in this channel is important on its own merits
because it confirms that the excess of tagged 6-jet events indeed
comes from top, or at least from a particle with a mass consistent
with that measured in the other decay modes.  Analysis of this final
state is not very likely to be feasible at the LHC.
 
In addition to an isolated high-$p_T$ electron or muon in the central
region of the detector, a single-lepton candidate event is required to
have at least four jets in order to perform a kinematical fit to the
top mass by a method similar to the one discussed above.  Here the
sample is much cleaner but still suffers from combinatorial
ambiguities in the reconstruction.  Including the two-fold ambiguity
in the neutrino $p_z$, it is four-fold if both $b$ jets are tagged,
12-fold if only one $b$ is tagged and 24-fold if none is.  Run 1
results in this channel are $173.3\pm 5.6 ({\rm stat.}) \pm 5.5 ({\rm
  syst.})$ [D\O~\cite{mt_D0_1lep}] and $176.1\pm 5.1 ({\rm stat.}) \pm
5.3 ({\rm syst.})$ [CDF~\cite{mt_CDF}].

It is interesting to note that even for the case when both $b$-jets
are tagged, MC simulations suggest that in only about half of the
cases does the best $\chi^2$ correspond to the correct matching of the
four leading jets to the appropriate quarks.  The other half are
roughly equally split between instances where all jets are matched to
partons, but the lowest $\chi^2$ did not choose the combination with
the correct assignments, and those where there are extra jets from
initial or final state radiation and the four leading partons from the
$t\bar {t}$ decay cannot be uniquely matched to the four leading jets
in the event.  At the LHC, $t\bar{t}$ events will have higher $p_T$,
on average.  This will often mean that the daughters of the two tops
will be on opposite sides of a plane.  Such hemispheric separation
will considerably alleviate these combinatorial problems.

A more recent analysis in the single-lepton channel by
D\O~\cite{mt_D0_1lep_new} makes a comparison of data with LO matrix
elements on an event-by-event basis, similar to that
suggested~\cite{mt.dilep.Kondo,Dalitz:wa} and used for the dilepton
channel discussed below.  This analysis requires the number of jets in
a candidate event to be exactly 4, and does not accord any special
status to events with $b$-tagged jets.  A likelihood function is
formed taking into account all possible permutations of jet
assignments, not just that with the lowest $\chi^2$.  The main
difference between this method and the previous is that that each
event now has an individual probability as a function of $m_t$.  This
probability, reflecting both signal and background, depends on all
measured variables in the event (except unclustered energy), with
well-measured events contributing more sharply to the extraction of
$m_t$ than those poorly measured.  The preliminary result, $m_t=179.9
\pm 3.6 ({\rm stat.}) \pm 6.0 ({\rm sys.})$~GeV, reflects a marked
reduction of the statistical uncertainty relative to the previous
result, which was based on the same data set but relied heavily on
explicit reconstruction of invariant masses.

Two alternatives to invariant mass reconstruction have been tried to
measure $m_t$ in the kinematically underconstrained dilepton channel,
$t\bar{t}\to \ell_1\nu_1 b\ell_2\nu_2\bar{b}$, which also suffers from
the smallest branching fraction.  In the first~\cite{mt.dilep.Kondo},
one hypothesizes a mass for the top quark, reconstructs the neutrino
momenta with a four-fold ambiguity for each lepton-$b$ pairing, and
calculates the probability of the final-state configuration to come
from a $t\bar{t}$ event of that $m_t$. For each event, a set of
assumed masses produces probability distributions to use as event
weights.  The preferred $m_t$ for an event can be taken as the maximum
or the mean of the distribution.  The distribution of preferred masses
for a set of candidate events is compared through a likelihood method
to the expected distribution from a combination of signal and
background, for given $m_t$.  As in the other channels, the central
value of the measurement is that with maximum likelihood.

Variants of this technique make use of more or fewer assumptions about
$t\bar{t}$ production details when obtaining the event probabilities.
For example, D\O\ has two different measurements, one using neutrino
kinematic distribution weights and another that uses production and
decay terms in the matrix element for the weights. The methods yielded
very consistent results.  The final result
is~\cite{mt_D0_2lep,Tollefson:wt} $m_t = 168.4\pm 12.3{\rm(stat.)}\pm
3.6{\rm(syst.)}$.

CDF's measurement in the dilepton channel used only information about
the expected pseudorapidity distributions of the neutrinos.  These
were chosen randomly from MC predictions, then the two neutrino
momenta were solved for. Each solution (ambiguity included) was
assigned a weight according to how well the derived \vmET matches that
measured. CDF's result is~\cite{mt_CDF,Tollefson:wt} $m_t = 167.4\pm
10.3{\rm(stat.)}\pm 4.8{\rm(syst.)}$.  CDF also used a likelihood fit
to kinematical variables that are sensitive to $m_t$: the $b$-jet
energy spectrum and the full event invariant mass~\cite{Abe:1997iz}.
Results from these are consistent, but suffer larger systematic
uncertainties.

The other method for the dilepton channel~\cite{Dalitz:wa} is based on
the observation that, modulo finite $W$ width effects, the $b$ quark
energy is fixed in the top quark rest frame. The top mass is then
given by
$m^2_t=\mean{m^2_{b\ell}}+\sqrt{M^4_W+4M^2_W\mean{m^2_{b\ell}}+
  \mean{m^2_{b\ell}}^2}$, where $\mean{m^2_{b\ell}}$ is the mean value
of $m^2_{b\ell}$ in the sample.  The results are generally consistent
with the likelihood methods.

The dilepton sample also contains a subsample of events that may prove
useful at the LHC for improving its uncertainties. Here, one looks for
events where one of the $b$ quarks hadronizes to $J/\Psi$, which
subsequently decays to $\ell^+\ell^-$, providing a cleaner and
more precisely measured sample. 
When the sister $W$ decays leptonically to $\ell^\prime\nu_{\ell^\prime}$,  
a strong correlation exists between $m_t$ and 
$m_{J/\Psi\,\ell^\prime}$~\cite{mt.dilep.Jpsi}.  
The top mass can be extracted essentially from the end point of the Gaussian
$m_{J/\Psi\,\ell^\prime}$ distribution.  In recent improvements to 
{\sc herwig}, matrix element corrections to radiative top decays are
known to cause a 1-1.5~GeV shift in the extracted
$m_t$~\cite{Corcella:2000wq}.  Study of this endpoint spectrum is
ongoing, and must take into account this MC improvement, to attain the
goal of 1~GeV precision in this channel.

\begin{figure}
  \epsfxsize=350pt \epsfbox{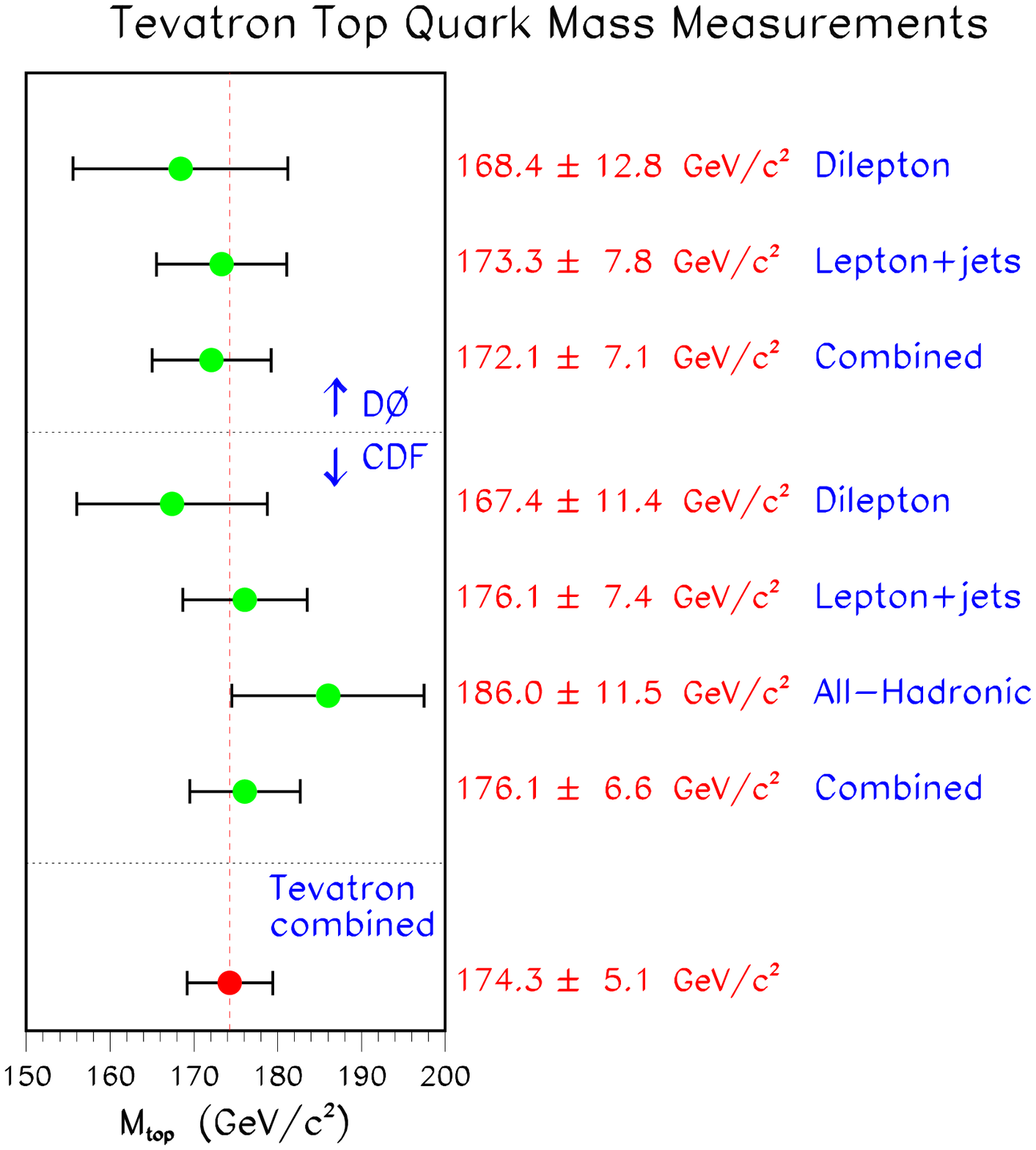}
\caption{Tevatron results for $m_t$ in the various channels, and
  the global average.}
\label{fig:tevmasses}
\end{figure}

The Tevatron average for $m_t$ is $174.3\pm 3.2 {\rm(stat.)}\pm 4.0
{\rm(syst.)}$~\cite{mt_D0_CDF}.  Fig.~\ref{fig:tevmasses} shows the
breakdown per channel, and the global average.
Table~\ref{tab:Tevatron_mass_err} summarizes the systematic uncertainties 
in the D\O\ and CDF Run 1 $m_t$ measurements in the various channels.  
As mentioned above, most of the systematic uncertainty comes from the 
jet energy scale.  
Experiments need to understand and maintain the calibration of
their calorimeters to high precision to help keep part of this
systematic under control.

With larger samples of events in Run 2 and at the LHC, both
statistical and systematic uncertainties will be reduced
significantly.  There are several reasons for this.  First, one can
afford to narrow the focus to samples with two $b$-tagged jets.  This
reduces combinatorics, but also energy scale uncertainty, since
specific energy corrections to $b$-jets can be applied to help with
the mass resolution. One can also choose specific subsets of events in
which, for example, the exact number of jets as expected from top are
found and in which the jet energies are particularly well measured (be
it fiducially or due to high energy).  Events with particular
topologies can similarly help. {\sc ATLAS} and {\sc CMS} plan to use
angular information and possible hemispheric separation of the two top
quarks as well, to assist with correct $b-W$ combination.
Additionally, with large integrated luminosity samples, the control
samples used to map the calorimeters' energy responses, such as
photon+jets and high-$E_T$ di-jets, will be less statistically limited
and will help reduce the jet energy scale uncertainty.

Another source of improvement in the mass measurement can come from a
better understanding of the treatment of initial and final state
radiation.  If the parton came from initial state radiation, including
it in the reconstruction would bias $m_t$ toward larger masses. If it
instead came from radiative top decay or the final state $b$ quark, it
must be included, else $m_t$ is measured to be too low. This issue has
been known for a long time, and addressed at the theoretical level
with exact calculations of the expected rates for and radiation
patterns of one additional hard parton~\cite{Orr97}. These authors
propose to assign additional hard jets in events to either production
or decay by calculating the following observables:
\begin{equation}
S_{prod} \; = \; | \, [ (p_{W^+}+p_b)^2 - m^2_t + im_t\Gamma_t]\times
              [ (p_{W^-}+p_{\bar{b}})^2 - m^2_t + im_t\Gamma_t] \, |
\end{equation}
\begin{equation}
S_1 \; = \; | \, [ (p_{W^+}+p_b)^2 - m^2_t + im_t\Gamma_t]\times
         [ (p_{W^-}+p_{\bar{b}}+p_j)^2 - m^2_t + im_t\Gamma_t] \, |
\end{equation}
\begin{equation}
S_2 \; = \; | \, [ (p_{W^+}+p_b+p_j)^2 - m^2_t + im_t\Gamma_t]\times
         [ (p_{W^-}+p_{\bar{b}})^2 - m^2_t + im_t\Gamma_t] \, |
\end{equation}
The extra jet is ``production'' if $S_{prod} < {\rm min}(S_1,S_2)$,
and ``decay'' otherwise; this assumes of course that in samples
containing hadronic $W$ decays, the correct assignment has already
been made for the $W$ jets (i.e. a radiative $W$ decay could be
identified).  How well the idea may apply under experimental
constraints remains to be evaluated.

\begin{table}[htb]
\def~{\hphantom{0}}
\caption{Channel-by-channel systematic uncertainties (GeV) in Tevatron Run 1 
top mass measurements.}
\label{tab:Tevatron_mass_err}
\vspace{3mm}
\begin{tabular}{|c|c|c|c|c|c|c|}
\toprule
\multicolumn{1}{|c}{Channel $\to$}&
\multicolumn{2}{|c}{Dilepton}&
\multicolumn{2}{|c}{Single lepton}&
\multicolumn{2}{|c|}{All-hadronic}\\
\colrule \hline
Systematic Category& CDF & D\O\ & CDF & D\O\ & CDF & D\O\ \\
\hline
Jet energy scale & 3.8 & 2.4 & 4.4 & 4.0 & 5.0 & ? \\
Model for Signal     & 2.8 & 1.7 & 2.6 & 1.9 & 1.8 & ? \\
MC generator    & 0.6 & 0.0 & 0.1 & 0.0 & 0.8 & ? \\
Uranium Noise/Multiple Interactions & 0.0 & 1.3 & 0.0 & 1.3 & 0.0 & ? \\
Model for Background & 0.3 & 1.0 & 1.3 & 2.5 & 1.7 & ? \\
Method for Mass Fitting    & 0.7 & 1.1 & 0.0  & 1.5 & 0.6  & ? \\\hline
Total            & 4.8 & 3.6 & 5.3 & 5.5 & 5.7 & ? \\\hline
\botrule
\end{tabular}
\label{table:Tev_mass}
\end{table}

\subsection{Spin}
\label{subsec:properties:spin}

All SM fermions have a left-handed weak gauge coupling, which mediates
their decays, if they decay. Only the top quark, because it is so
massive, decays before it hadronizes or its spin flips, thus leaving
an imprint of its spin on its angular decay distributions. But how do
we even know that the top quark candidate is a fermion? First, if it
were spin 0 or 1, we would have to postulate an additional unobserved
daughter to conserve overall spin. Furthermore, although Tevatron and
LHC use unpolarized beams and therefore produce unpolarized top quark
pairs, for spin 0 their spins would be uncorrelated, whereas for spin
1 they would be, although this correlation has not been considered.
The spin correlations arising from a spin 3/2 scenario have also not
been considered. However, a simple argument against spin 3/2 is that
the $t\bar{t}$ cross section would be much larger. This was in fact
how the tau lepton was determined to be spin 1/2.

As a spin 1/2 fermion, the SM top quark has decay angular
distributions $d\Gamma / d(\cos\theta^*_i) \propto 1 + \alpha_i
\cos\theta^*_i$, where $\theta^*_i$ is the angle of decay particle $i$
in the top quark rest frame with respect to the top quark spin
($i=\ell^+,\nu,b$, or $\bar{d},u,b$), and $\alpha_i$ is the {\it spin
  analyzing power} of particle $i$. At LO, $\alpha_i = 1,-0.32,-0.41$
($\alpha_i$ have opposite signs for top quark and anti-top quark),
making the outgoing charged lepton or down-type quark not tagged as a
$b$ the ideal spin correlation analyzer. If one uses the down-type
quark in hadronic $W$ decays, the QCD NLO corrected value must be
used~\cite{Brandenburg:2002xr}: $\alpha_{\bar{d}} \simeq 0.93$. For
top quark pair production, because the spins are correlated, one plots
a double differential distribution~\cite{Stelzer:1995gc,Mahlon.Parke},
\begin{equation}
{1\over\sigma} \,
{d^2\sigma\over d(\cos\theta_i)d(\cos\theta_{\bar{i}})}
\; = \;
{1\over 4} (1 \, - \, C \, \alpha_i \, \alpha_{\bar{i}}
                   \, \cos\theta_i \, \cos\theta_{\bar{i}} )
\; ,
\end{equation}
where $\theta_i(\theta_{\bar{i}})$ is now the angle of the
$i^{th}(\bar{i}^{th})$ decay product with respect to the chosen spin
axis in the top (anti-top) quark rest frame; and $C$ is the {\it spin
  correlation coefficient} - the relative fraction of like-spin top
quarks produced, in the spin basis considered. Near threshold,
$t\bar{t}$ produced by quark pairs is in a $^3S_1$ state, whereas
gluon production yields a $^1S_0$ state, so the two components will
have different spin correlations, $C_{q\bar{q}}$ and $C_{gg}$.
Observing the overall correlation would confirm that the top quark is
indeed the SM partner of the bottom quark with a left-handed weak
coupling.

\begin{figure}
  \epsfxsize=240pt \epsfbox{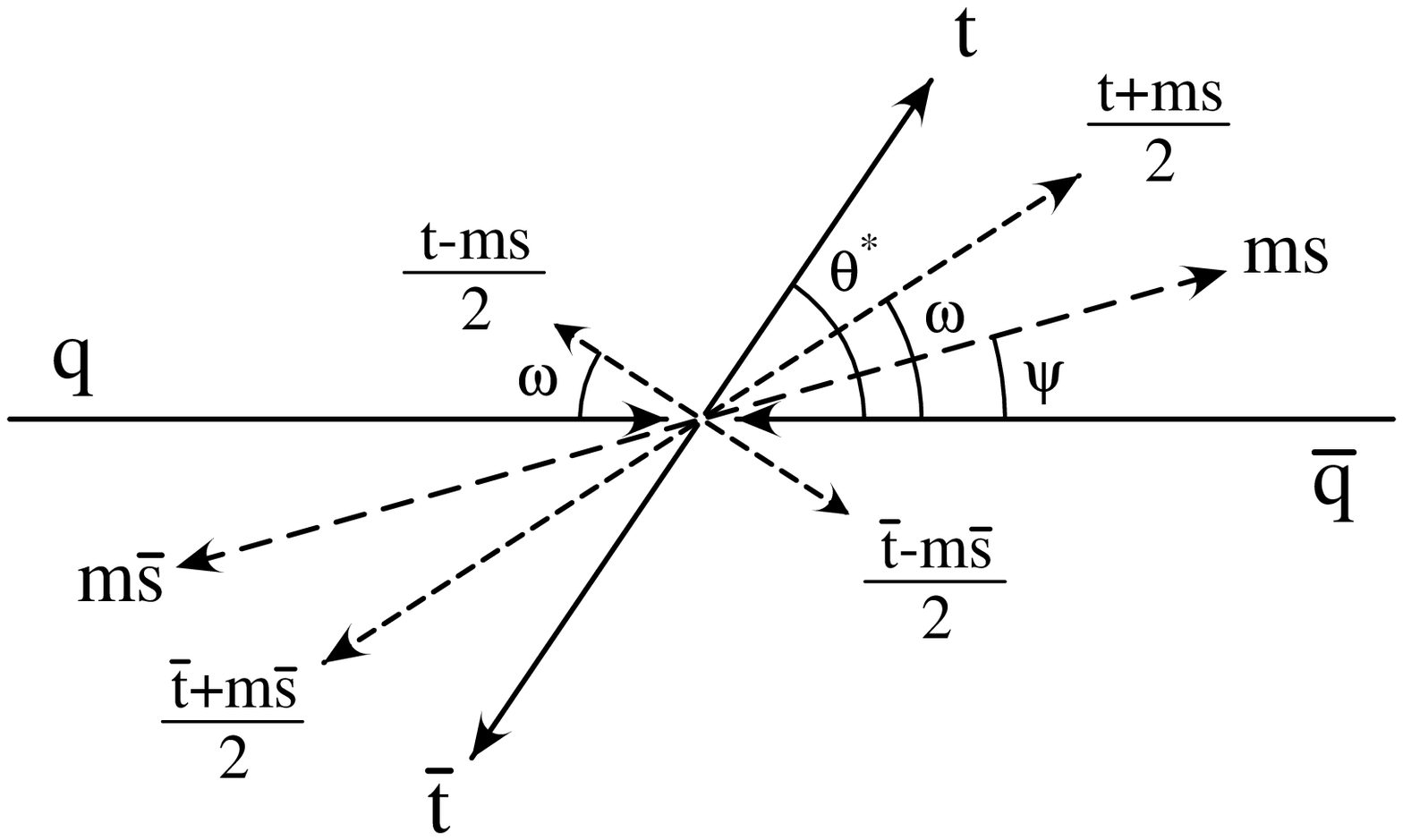}
  \caption{$t\bar{t}$ rest frame (``zero momentum frame'') for
    $q\bar{q}\to t\bar{t}$ at hadron colliders, from
    Ref.~\cite{Mahlon.Parke}(b). $t(\bar{t})$ are the (anti-)top quark
    momenta, $s(\bar{s})$ are the (anti-)top quark spin vectors.
    $\theta^*$ is the flight direction of the top quark, $\psi$ is the
    direction of the off-diagonal spin bases, and $\omega$ is the
    preferred emission direction of the down-type fermion in top quark
    decay for up-down ($t+ms$) and down-up ($t-ms$) spin
    configurations.  All angles are with respect to the $p$ beam
    direction.}
  \label{fig:spinbasis}
\end{figure}

The overall spin correlation coefficient $C$ varies strongly depending
on spin basis and which initial state parton type dominates. Because
$t\bar{t}$ production at the Tevatron is predominately
quark-initiated, while at the LHC it arises mostly from initial
gluons, different spin bases optimize analyses for the two machines.
At the Tevatron this is the ``off-diagonal'' basis of
Ref.~\cite{Mahlon.Parke}, where the spin basis angle $\psi$ with
respect to the proton beam direction is a function of the speed and
production angle $\theta_t$ of the top quark with respect to the
incoming $p$ direction in the zero momentum frame (ZMF):
\begin{equation}
\tan\psi \; = \; 
{\beta^2 \, \sin\theta_t \, \cos\theta_t \over
 1 \, - \, \beta^2 \, \sin^2\theta_t           } \; .
\end{equation}
This basis is illustrated in
Fig.~\ref{fig:spinbasis}~\cite{Mahlon.Parke}. At the LHC, the
``helicity basis'' is optimal, which resolves spin along the flight
direction of the top quarks in the ZMF. The NLO corrections to $C$ are
known to be ${\cal O}(10\%)$, and so will not greatly affect an
analysis~\cite{spin.NLO}. However, the uncertainty in $C$ even at NLO
is unexpectedly large at the Tevatron. Because $C_{gg}$ contributes
with opposite sign to $C_{q\bar{q}}$, the overall value is quite
sensitive to uncertainties in the gluon structure function at high
$x$. Thorough study of PDF uncertainties will be required to resolve
this. It is not as serious an issue at the LHC, as this process probes
$g(x)$ at low $x$, where the PDF uncertainties are quite small, and in
any case the scale uncertainty at NLO dominates over PDF uncertainties
for this machine. At the Tevatron in the off-diagonal basis,
$C_{NLO}=0.806^{+2.9\%}_{-4.0\%}(\mu)^{+4.0\%}_{-8.9\%}({\rm PDF})$,
and in the helicity basis at the LHC,
$C_{NLO}=0.311^{+6.4\%}_{-10.6\%}(\mu)^{+6.8\%}_{-0.0\%}({\rm
  PDF})$~\cite{spin.NLO}.

Because the spin analyzing power of the charged lepton (leptonic
decay) or $d$ quark (hadronic decay) is maximal, they are the natural
choice for observing the correlations. The dilepton $t\bar{t}$ sample
has the least background contamination, but because of the two missing
neutrinos can be reconstructed only statistically. Flavor tagging is
not possible among the light quarks, but the down-type quark is
typically the least-energetic quark in $W$ decay in the top quark rest
frame. In principle, then, use of the single-lepton and all-hadronic
channels is possible, but needs further investigation.

If the top quarks decay isotropically, then $C=0$ (no correlation).
New physics such as $CP$ violation or a right-handed coupling
component would also alter the predicted value of
$C$~\cite{Kane:1991bg}. The task then is to determine the achievable
level of uncertainty on $C$ at Tevatron and LHC. {\sc D\O\ } has
performed an analysis of their dilepton samples~\cite{Abbott:2000dt}.
While the statistics were too poor to give a strong result, they
clearly established that the measurement can be performed. Run 2
expectations are that $C=0$ can be ruled out at better than the
$2\sigma$ level with 2~fb$^{-1}$ of data. At the LHC, {\sc CMSJET}
simulation~\cite{Beneke:2000hk} estimates a measurement of $C = 0.331
\pm 0.023$ (statistical errors only, LO simulation) for the SM, more
than good enough to rule out the isotropic decay case. Polarimetry of
the $b$ quark has been proposed to enhance spin correlation
analyses~\cite{Nelson:2001su}, but has not yet been investigated by
the experimenters. Of course, the ultra-low background environment,
beam polarization, and $\sqrt{s}$ tuning of a LC would be ideal for
precision spin and spin correlation measurements~\cite{Parke:1996pr}.

Because all three modes of single top quark production
(Fig.~\ref{fig:singletop}) can be observed at both Tevatron and LHC,
it is useful to consider spin for these cases as well.  Here the
interesting distribution is the angle $\theta$ between the charged
lepton in the top quark decay and the chosen spin
axis~\cite{t.spin.theory,Boos:2002xw}:
\begin{equation}
{1\over\sigma_T} \, {d\sigma\over d\cos\theta} \; = \;
{1\over 2} \biggl[ 1 + C'\cos\theta \biggr] \; , \; \; 
C' = {N_\uparrow - N_\downarrow \over N_\uparrow + N_\downarrow} \, .
\end{equation}
where $N_{\uparrow (\downarrow)}$ is the number of top quark events
produced spin up (down) in the frame considered. The spin asymmetry
$C'$ in this case is maximized by choosing the spin basis that most
strongly correlates with the down-type quark on the production side.
For $W^*$ production, this is simply the antiproton direction at the
Tevatron~\cite{t.spin.theory}. The $Wg$-fusion process is more
challenging due to NLO complications in the initial and final states
as the ZMF cannot be defined. Here one optimally chooses the
``$\eta$-beamline'' basis, which is defined as the beamline most
closely aligned with the forward scattered quark that supplied the
fusing $W$~\cite{t.spin.theory}. For $Wt$ production the ideal basis
is defined by the down-type fermion from both $W$
decays~\cite{Boos:2002xw}. This channel has severe experimental
problems reconstructing the top quark rest frame for most decay
channels, but is under investigation.

One study~\cite{Stelzer:1995gc} noted this is also a crucial test of
the CKM matrix element $V_{tb}$: since $\Gamma_t$ is nearly
proportional to $|V_{tb}|^2$,\footnote{assuming $|V_{tq}|\ll|V_{tb}|$
  for $q=d,s$.} if $V_{tb}$ were small due to a fourth generation,
then top would decay on average after the spin flip time
$m_t/\Lambda^2_{QCD}$ - the spin correlation would not be seen!  This
provides the constraint $|V_{tb}|> 0.03$.

\subsection{Charge}
\label{subsec:properties:charge}

The electric charge of the top quark has not actually been measured.
While it is not widely supposed that its value is not that of the SM,
there do exist exotic theories where the top quark is actually much
heavier, and the Run 1 observation is of another exotic quark of
charge $Q = -4/3$~\cite{Chang:1998pt}. Techniques to measure this
directly at hadron colliders have been explored using the sample of
single lepton events that contain a hard photon~\cite{Baur:2001si}:
$t\bar{t}\to \gamma\ell\nu b jj\bar{b}$ ($j$ is a jet from $W\to
q_1\bar{q}_2$).

The photon can be radiated from any electrically charged particle in
the process, which means that contributions arise from radiation in
top production (including quark initial states), radiative top decay,
and radiative $W$ decay. The contribution of radiative $W$ decay is
SM-like and its influence can be removed by requiring that the
invariant mass of the $jj\gamma$ system and the transverse mass of the
$\ell\gamma\vmpT$ system be larger than 90~GeV. Events are dominated
by photons produced in top production if one imposes the cuts:
\begin{equation}
\label{eq:ttgamcuts}
m(b_{1,2}jj\gamma) \; > \; 190 \; {\rm GeV} \; , \; \;
m_T(b_{2,1}\ell\gamma\vmpT) \; > \; 190 \; {\rm GeV} \; .
\end{equation}
At Tevatron energies, photon radiation from the initial state quark
pairs (which constitutes about $90\%$ of $t\bar{t}$ events) dominates
the cross section, so $Q_t = -4/3$ increases the cross section of this
sample by only about $20\%$. At the LHC, however, where $gg\to
t\bar{t}$ dominates, it is enhanced by a factor 2.6, since the cross
section is roughly proportional to $Q^2_t$. Radiative decay samples
are selected by selectively changing one of the relative symbols for
the cuts of Eq.~\ref{eq:ttgamcuts}. In these cases, the sample cross
sections actually decrease if $Q_t = -4/3$, due to interference
between radiation from the $t$, $W$ and $b$ lines.

More useful is to examine the $p_T$ and angular distributions of
photons for the three $t\bar{t}\gamma$ samples, which are anomalous in
the case of exotic charge assignment. For example the photon is
typically closer to the lower-energy $b$ quark. The distributions can
be used to perform a $\chi^2$ test to distinguish the $Q_t =
+2/3,-4/3$ hypotheses. $Q_t$ for this purpose is treated in the
literature as a continuous quantity, rather than discrete, because the
strict requirement of a viable EW model is simply that the two
partners of an $SU(2)$ doublet differ by one unit of charge.  However,
the models that allow for this realization are quite strange, so we
choose to present results in terms of distinctly separating the two
discrete charge assignments.  Estimates are that Tevatron Run 2 could
confirm $Q_t = +2/3$ at $95\%$~C.L. with about 20~fb$^{-1}$ of data
using the photon distributions, while the LHC could do this at
$100\%$~C.L. with 10~fb$^{-1}$.  A 500~GeV LC could achieve this as
well with ${\cal O}(100)$~fb$^{-1}$ of data~\cite{Abe:2001nq}.

Alternatively, one can look for a few very clean single lepton
$t\bar{t}$ events where either the $b$ jet charge is measured, or the
$b$ from the leptonic top decay decays
semi-leptonically~\cite{Baur:2001si}.  Since $Q_t = Q_b + Q_{\ell}$,
the latter could work even at the Tevatron if experiments are lucky to
see a few clean such events.  However, measuring $b$ jet charge is
less well explored.

\subsection{Gauge couplings}
\label{subsec:properties:coups}

We know via observation of $p\bar{p}\to t\bar{t}\to b\bar{b}W^+W^-$ at
the expected SM rate, and non-observation of other decays (including
radiative QED), that the top quark gauge couplings to
$g,W^\pm,Z,\gamma$ are roughly SM-like. These must now be measured
precisely; anomalous coupling analyses are the most appropriate. CP
violation in the top sector is normally addressed in this language,
via the CP-even and -odd terms in the effective Langrangians used.

The motivation for studying anomalous QCD top quark gauge couplings is
that they naturally arise in dynamical EW symmetry breaking models
such as technicolor or topcolor. They have been explored for the
Tevatron~\cite{topgaugecoups.TeV,Martinez:2001qs,Hikasa:1998wx} and
LHC~\cite{Beneke:2000hk,Martinez:2001qs} (see also references
therein). The effective Lagrangian appears as the SM term plus
chromoelectric and chromomagnetic dipole moment terms,
\begin{equation}
{\cal L}_{t\bar{t}g} \; = \; \bar{t} \, \biggl[ \;
- g_s \gamma^\mu G_\mu \; 
- i {g_s \hat{d}_t'\over 2m_t} \sigma^{\mu\nu} \gamma_5 G_{\mu\nu} \;
- {g_s \hat{\mu}_t'\over 2m_t} \sigma^{\mu\nu} G_{\mu\nu} \;
\biggr] \, t \; .
\end{equation}
Both terms flip chirality; the chromomagnetic moment $\hat{\mu}_t'$ is
CP-even, and the chromoelectric moment $\hat{d}_t'$ is CP-odd,
enabling use of CP-even and -odd observables to separate their
effects. Because the CP-even chromomagnetic moment interferes with the
SM vertex, observables are potentially sensitive to the sign of the
coupling. One calculational detail is that for $gg\to t\bar{t}$
subprocesses, an additional dimension-5 operator must be introduced to
preserve gauge invariance, corresponding to an effective $ggt\bar{t}$
4-point interaction. There is also a SM loop contribution to the
chromomagnatic moment, which depends on the Higgs boson mass. For
example, for $M_H = 100$~GeV, this leads to a $2.5\%$ correction to
$\sigma_{t\bar{t}}$ at the LHC, which is smaller than the expected
measurement uncertainty~\cite{Martinez:2001qs}. The same study shows
${\cal O}(10-20)\%$ changes can occur in models containing two Higgs
doublets or additional matter content, such as the MSSM.

Unfortunately, Tevatron studies have shown that these moments lead
mostly to overall $t\bar{t}$ rate changes, due to threshold effects
dominating the angular distributions. Only for very large values of
$d_t',\mu_t'$ might one expect to observe shape changes in such
distributions as the top quark emission angle in the center-of-mass
frame, or for dileptonic decays at the Tevatron,
\begin{equation}
\hat{O}_L \; = \; 
{1\over m^3_t |P|^2} \, P\cdot (Q_+\times Q_-)\, P\cdot (Q_+ - Q_-) \; , 
\end{equation}
where $P(Q_+,Q_-)$ is the momentum vector of the
proton($\ell^+,\ell^-)$, also in the CM frame.  Even then, the
statistics at Run 2 may not be sufficient to explore this with
confidence. Furthermore, constraints from $b\to s\gamma$ on the
chromomagnetic moment are already an order of magnitude better than is
achievable at Tevatron~\cite{Martinez:2001qs}. The prospect for
explicit CP-odd observables for the chromoelectric moment is greater,
but further study with detector simulation and up-to-date Run 2
expectations is needed. Unfortunately, the literature on $t\bar{t}g$
anomalous couplings contains a wide variety of conventions. For LHC
studies this is particularly noticeable: results are extremely
difficult to compare, both with each other and with other experimental
constraints such as from $b\to s\gamma$. This should be rectified in
the near future, to clarify what exactly can be learned.

At hadron colliders, anomalous $t\bar{t}\gamma$ and $t\bar{t}Z$
couplings can be explored only via associated production, as EW
s-channel contributions to top pairs are far too suppressed relative
to QCD. Up-to-date predictions for these SM rates may be found in
Refs.~\cite{Baur:2001si,Maltoni:2002jr}. No anomalous coupling
analysis for these cases has yet been performed, beyond the top charge
measurement of the former. At a LC, these can be studied in direct
$t\bar{t}$ production quite
precisely~\cite{Abe:2001nq,Aguilar-Saavedra:2001rg}.

Anomalous $t\bar{b}W$ couplings have been explored for hadron
colliders in the context of $t\bar{t}$ production and
decay~\cite{Kane:1991bg}, and more recently of single top
production~\cite{Boos:1999dd,Hikasa:1998wx,Beneke:2000hk}. For
$t\bar{t}$ production the previously discussed limit on right-handed
$W$ bosons in top decay is part of this subject, but not normally
discussed in anomalous coupling language. The effective Lagrangian is
\begin{equation}
{\cal L} \; = \; {gV_{tb}\over\sqrt{2}} \biggl[
W^-_\mu \, \bar{b} \, \gamma_\mu P_- \, t \, - \,
{1\over 2M_W} W^-_{\mu\nu} \, \bar{b} \, \sigma^{\mu\nu} 
(F^L_2 P_- + F^R_2 P_+) \, t
\biggr] \; + \; h.c.
\end{equation}
where $W^\pm_{\mu\nu}$ is the field strength tensor and $P_\pm =
(1\pm\gamma_5)/2$; $F^{L,R}_2 = 0$ in the SM. The non-SM term is
proportional to the particle momentum, and is realized by an anomalous
contribution to the cross section at high $p_T$. In practice one uses
the $W$, $b$, or $bb$ systems, depending on which single top
production component is isolated. Even with 2~fb$^{-1}$ at the
Tevatron, limits of approximately $-0.18 < F^L_2 < +0.55$ and $-0.24 <
F^R_2 < +0.25$ could be achieved, assuming a $10\%$ systematic
uncertainty. At the LHC this would improve by a factor of 2-3. It is
important that this theoretical study be followed up by one with
detector simulation to include systematic uncertainties, which will
likely be limiting. Limits from a LC would be better by up to an order
of magnitude. As a final note, Ref.~\cite{Larios:1999au} pointed out
that CLEO data on $b\to s\gamma$ is already more constraining on
right-handed $tbW$ couplings than would be achievable at any planned
future colliders.

\subsection{Lifetime and $V_{tb}$}
\label{subsec:properties:lifetime}

The CKM matrix element $V_{tb}$ is intimately related to the top quark
lifetime, so it is natural to discuss them together, even though they
are often treated as separate topics. We usually speak of the
lifetimes of quarks (charm and bottom) and leptons (muon and tau),
rather than their intrinsic widths, because they are some fraction of
a second that is measureable in the laboratory. Indeed it is such
``long'' lifetimes that allow high resolution vertex detectors to see
the displaced decay vertices of $\tau$ leptons, $b$ and $c$ quarks in
collider experiments. Like the other fermions, top decays only weakly,
so does it also have a long life?  Fortunately, no! The top quark
lives about $4\times 10^{-25}$~s, almost an order of magnitude more
fleeting than the time it takes for a colored particle to hadronize.

A particle's lifetime is the inverse of its decay width,
$\tau={\hbar\over\Gamma}$. In fact we calculated the top lifetime by
first calculating its decay width. For extremely short-lived states,
it's more useful to discuss the width, rather than the lifetime.
Ignoring the $b$ quark mass, at LO the top quark $bW$ partial width is
\begin{equation}
\label{eq:twidth}
\Gamma \, (t\to Wb) \; = \; 
{G_F\over 8\pi\sqrt{2}} \, m^3_t \, |V_{tb}|^2 \, 
\biggl(
1 \, - \, 3\,{M^4_W\over m^4_t} \, + \, 2\,{M^6_W\over m^6_t}
\biggr) \; = \; 1.56 \, {\rm GeV} \; .
\end{equation}
The NLO result is 1.42~GeV~\cite{width.NLO}. Note that the NLO value
cannot be used in a LO matrix element calculation - it will give the
wrong ${\cal B}(t\to bW)$, because the other couplings are at LO! This
partial width is proportional to $|V_{tb}|^2$, just as the other SM
decays, $t\to sW,dW$, are proportional to $|V_{ts}|^2,|V_{td}|^2$,
respectively.  These are a $\approx 0.2\%$ correction to the total
width, $\Gamma_t = \sum_q\Gamma_{tq}$, if there are indeed only 3
generations of quarks, for which case $0.9990 < |V_{tb}| < 0.9993$. We
can be confident that $|V_{tb}| \gg |V_{ts}|,|V_{td}|$ even without
the low energy unitarity constraints, from the CDF
measurement~\cite{Affolder:2000xb}
\begin{equation}
{{\cal B}(t\to bW) \over {\cal B}(t\to qW)} \; = \; 
{|V_{tb}|^2 \over |V_{tb}|^2 + |V_{ts}|^2 + |V_{td}|^2} \; = \; 
0.94^{+0.31}_{-0.24} \; ,
\end{equation}
which looks for the fraction of tagged $b$ jets in $t\bar{t}$ decays.

It is interesting to consider what happens if there are more than
three generations, in which case unitarity constraints on $V_{tb}$
from low energy data are virtually meaningless. From EW precision data
we know the rho parameter quite precisely. For four generations its
value is~\cite{Willenbrock:1997nr}
\begin{equation}
\rho \; \simeq \; 
1 \, + \, {3G_F\over 8\sqrt{2}\pi^2} 
\biggl[ m^2_t|V_{tb}|^2 + m^2_{t'}|V_{t'b}|^2 \biggr] \; = \;
1 \, + \, {3G_F\over 8\sqrt{2}\pi^2} 
\biggl[ m^2_t + \epsilon^2 (m^2_{t'}-m^2_t) \biggr] ,
\end{equation}
where $t'$ is the up-type fourth-generation quark, and unitarity in
the fourth generation requires that $|V_{tb}|^2 = 1-\epsilon^2$,
$|V_{t'b}|^2 = \epsilon^2$ (given our belief in very small
$V_{ts},V_{td}$). It is obvious that either $\epsilon$ is small or the
top quark and the fourth generation up-type quark are nearly
degenerate. The latter case would be discovered quite soon, the fourth
generation issue is not one of great concern.

For unstable particles, the width exhibits itself as a spread in the
invariant mass distribution of the decay products, the Breit-Wigner
lineshape. Unfortunately, the top quark width is narrower than
experimental resolution at a hadron collider, so neither Tevatron nor
LHC will be able to determine this directly. (One can set limits of
the detector resolution, but this will never be competitive with
${\cal B}$ checks and other methods.) But determining it is not
impossible: one resorts instead to an indirect method of combining
several other results which depends on $\Gamma_t$. This requires
observation of both $t\bar{t}$ and single-top production (in at least
one of the three channels) and some mild theoretical assumptions that
can be checked, within limits, via detailed studies of decay angular
distributions.  One has to assume that QCD governs the $t\bar{t}$
production and that the $t\bar{b}W$ vertex is the standard $SU(2)_L$
weak gauge vertex; both are eminently reasonable, and can be checked
via anomalous couplings analyses we discussed earlier, which look for
deviations in various differential distributions and so are not
reliant on only the total rate. All the necessary cross sections are
known at NLO or better.

The measurement is linked to $|V_{tb}|$, discussed previously.  First,
measure $\sigma_{t\bar{t}}\times ({\cal B}(t\to bW))^2$; given trust
in QCD and the NLO+NNLL rates, this yields ${\cal B}(t\to bW)$ to
$5\%$ at Tevatron Run 2 and $3\%$ at the LHC. Second, measure the SM
rate of single-top production, which is really $\sigma_{tX}\times
{\cal B}(t\to bW)$.  The production cross section, which is
proportional to the partial width $\Gamma (t\to bW)$, is obtained by
dividing out the known ${\cal B}$.  This is really a measurement of
$g_W\times|V_{tb}|$. Assuming exact dependence on the SM gauge
coupling $g_W$, this directly determines $|V_{tb}|$ - to $12\%$ at the
Tevatron (2~fb$^{-1}$), and $5\%$ at the LHC, where the measurement
will be systematics limited. The top quark total width is then the
partial width, given by Eq.~\ref{eq:twidth}, divided by ${\cal B}$.
Precision will be similar to that for the partial width to $bW$.

For the total width measurement it is expected that the
three-generation value of $|V_{tb}|$ would be used, as it is known
much more precisely from low energy data than can be measured
directly. The technique to measure $|V_{tb}|$ directly at hadron
colliders simply establishes to a high degree of confidence that no
fourth generation exists, which is already highly disfavored by EW
precision data. One may also cross-check ${\cal B}(t\to bW)$ by taking
the ratio of dilepton to single lepton events in $t\bar{t}$
production.

\subsection{Yukawa coupling}
\label{subsec:properties:yukawa}

Yukawa couplings relate the matter content of the SM to the source of
mass generation, the Higgs sector. For the top quark in the SM this is
written as a Lagrangian term ${\cal L} \, = \, -Y_t \bar{t}_L \phi t_R
\; + h.c.$. When the Higgs field $\phi$ acquires a vacuum expectation
value (vev) $v$, $\phi \to {1\over\sqrt{2}}(v + H)$, the vev term becomes
the mass and and the field term $-{1\over\sqrt{2}}Y_t\,\bar{t}_LHt_R$
becomes the interaction of a pair of top quarks with the physical
Higgs boson. Thus, the top quark mass is fundamentally related to the
Higgs vev and its Yukawa coupling,
$m_t\,=\,{Y_t\,v\over\sqrt{2}}$. Since $v=246$~GeV and
$m_t=174.3$~GeV, it appears that $Y_t$ is exactly 1, a theoretically
interesting value, leading to speculation that important new physics
may be accessed via top quark studies. The task then is to verify
this, by probing the Higgs-top interaction and therefore the mechanism
of fermion mass generation. This turns out to be the most difficult
top quark property to measure!

There are three methods to consider at hadron colliders: inclusive
Higgs production $gg\to H$, mediated dominantly by a top quark loop;
or associated production with a single top quark, or a pair.  Of
these, $gg\to H$ has the largest cross section, but is only minimally
useful. First, there is the possibility that additional undiscovered
particles mediate a loop contribution, which may not be
separable. Second, in 2HDM scenarios, the bottom quark contribution
introduces an additional uncertainty since it must be separated. While
this channel is still useful, direct access to $Y_t$ via top quark
associated production is more attractive.

One would expect the cross section for $tH$ production to be larger
than that for $t\bar{t}H$, which is more than two orders of magnitude
smaller than $gg\to H$ due to phase space suppression, since there is
more phase space available with only one top quark.  Unfortunately,
this is not the case, due to a unitarity cancellation between $tH$
diagrams~\cite{Maltoni:2001hu}, rendering this channel useless. It was
hoped that at the Tevatron $t\bar{t}H;H\to b\bar{b}$ could be observed
for a light Higgs, due to the highly unique final
state~\cite{Goldstein:2000bp}. However, the unexpectedly large,
negative QCD NLO corrections~\cite{ttH.NLO} have all but quashed this
hope. At the LHC $t\bar{t}H;H\to b\bar{b}$ is probably visible for a
very light Higgs~\cite{ttH.H2bb.LHC}, and it would be possible to
observe $t\bar{t}H;H\to W^+W^-$ for Higgs masses larger than about
120~GeV~\cite{Maltoni:2002jr,ttH.H2WW}. The statistical uncertainty on
$Y_t$ for the latter could be as small as $10\%$, but the systematics
have not been estimated.

At hadron colliders, simply measuring any of these production rates
is not sufficient to measure $Y_t$, despite the commonly held belief
that $t\bar{t}H$ grants ``direct access'' to the top Yukawa
coupling. The cross section is a convolution of $Y_t$ and the Higgs
branching ratio, which is a priori unknown. Only by multiple Higgs
measurements that determine all the Higgs branching ratios can such a
cross section measurement be meaningful. Thus, this aspect of top
quark physics is inextricably linked to Higgs physics. At the LHC,
where a Higgs signal would not be so statistically limited and would
appear in multiple channels, branching ratios can be determined
indirectly with mild theoretical assumptions~\cite{Zeppenfeld:2000td},
making interpretation of the rates useful. However, an unbiased
measurement of $Y_t$ will almost certainly require additional Higgs
data from a LC. There is an important corollary to this, for the case
of a large excess of events: even if the branching ratio to the
observed final state is assumed to be unity, strong constraints can be
put on models where $Y_t$ is significantly enhanced over SM
expectations. This can happen e.g. in topcolor assisted technicolor
models~\cite{Leibovich:2001ev}.

\section{Summary}
\label{sec:summary}

Discovery of the top quark has opened up a rich field of physics that
is justifiably attracting much attention.  Careful examination of its
production and decay characteristics, and precision measurement of its
mass and other properties, are needed to test the SM.  Theoretical and
experimental efforts towards must proceed hand-in-hand to this end.
The top quark may itself lead to the discovery of new physics: the
large top mass may well indicate a special role in electroweak and
flavor symmetry breakings, and particles yet unobserved may show up in
the production or decay of the top.  It is also important to
understand top quark events as fully as possible, because they will
constitute a strong background to many potential new physics signals
in other searches.

For the next 5 years or so, direct studies of the top quark belongs to
the ongoing Run 2 of the Tevatron.  While collider upgrades have
resulted in higher rate of production through increases in energy
(resulting in a cross-section enhancement of about $40\%$ for pairs
and $60\%$ for single top, compared to Run 1) and integrated
luminosity (50 times or more), detector upgrades will allow superior
background suppression.  We expect that data samples containing ${\cal
  O}(100)$ times as many top quarks as presently available will be
collected during this period.  After that, the LHC will dominate the
field, delivering another factor of ${\cal O}(100)$ increase in top
quark yield.  Better understanding of QCD dynamics is required to make
full use of the rich statistics of top events at hadron colliders,
leaving plenty of room for work to prepare for the LHC era.  High
energy physicists around the world have started planning for a future
$e^+e^-$ linear collider, which may become operational around 2015.
Such a machine will offer new means for precision studies of the top
quark properties and dynamics.

In closing, we quote an observant colleague~\cite{Top-ology}, ``In
physics, one discovery often leads to others.  Top opens a new world
-- the domain of a very heavy fermion -- in which the strange and
wonderful may greet us.''

\section*{Acknowledgement}

The authors would like to thank the following people for invaluable
advice: Jerry Blazey, Arnd Brandenburg, Tom Ferbel, Mark Kruse, Eric
Laenen, Michelangelo Mangano, Steve Martin, Carlo Oleari, Lynne Orr,
Stephen Parke, John Parsons, Rob Roser, Zack Sullivan, and Scott Willenbrock. 
DC's work was supported in part by a grant from the U.S.  National Science
Foundation, and JK's by the U.S. Department of Energy.  DR would like to
thank Fermilab and DESY, where parts of this work were completed.

\end{document}